\RequirePackage{fix-cm}
\documentclass[smallextended]{svjour3}
\smartqed  
\usepackage{appendix}
\usepackage{amsmath}
\usepackage{graphicx}
\usepackage{lineno}
\usepackage{array}
\usepackage{longtable}
\usepackage{natbib}
\usepackage{color}
\usepackage{float}
\usepackage{comment}
\usepackage{physics}
\usepackage{esint}
\usepackage[colorlinks,linkcolor=blue,bookmarksopen,bookmarksnumbered,citecolor=blue,urlcolor=blue]{hyperref}
\setcitestyle{aysep={}}

\renewcommand{\d}{\textnormal{d}}

\renewcommand{\vec}[1]{\boldsymbol{#1}}
\renewcommand{\perp}{h}

\newcommand{\spav}[1]{\left\langle {\smash{#1}} \right\rangle_{xy}}
\newcommand{\iav}[1]{\left\langle {\smash{#1}} \right\rangle{^f}}
\newcommand{\sav}[1]{\left\langle {\smash{#1}} \right\rangle}
\newcommand{\volav}[1]{\sav{#1}}

\newcommand{\review}[1]{\textcolor{black}{#1}}
\newcommand{\U}[1]{\textnormal{#1}}

\newcommand*\patchAmsMathEnvironmentForLineno[1]{%
\expandafter\let\csname old#1\expandafter\endcsname\csname #1\endcsname
\expandafter\let\csname oldend#1\expandafter\endcsname\csname end#1\endcsname
\renewenvironment{#1}%
{\linenomath\csname old#1\endcsname}%
{\csname oldend#1\endcsname\endlinenomath}}%
\newcommand*\patchBothAmsMathEnvironmentsForLineno[1]{%
\patchAmsMathEnvironmentForLineno{#1}%
\patchAmsMathEnvironmentForLineno{#1*}}%
\AtBeginDocument{%
\patchBothAmsMathEnvironmentsForLineno{equation}%
\patchBothAmsMathEnvironmentsForLineno{align}%
\patchBothAmsMathEnvironmentsForLineno{flalign}%
\patchBothAmsMathEnvironmentsForLineno{alignat}%
\patchBothAmsMathEnvironmentsForLineno{gather}%
\patchBothAmsMathEnvironmentsForLineno{multline}%
}

\begin{document}

\title{Multi-scale analysis of flow over heterogeneous urban environments}

\author{Maarten van Reeuwijk \and
        Jingzi Huang
}

\institute{      
           Maarten van Reeuwijk \at 
           Department of Civil and Environmental Engineering, Imperial College London, London SW7 2AZ, UK 
\email{m.vanreeuwijk@imperial.ac.uk} \\
           Jingzi Huang \at 
           Department of Civil and Environmental Engineering, Imperial College London, London SW7 2AZ, UK 
\email{jingzi.huang17@imperial.ac.uk}
}

\date{Received: DD Month YEAR / Accepted: DD Month YEAR}

\maketitle

\begin{abstract}
A \review{computationally efficient} multi-scale planar-averaging framework for urban areas is developed, which enables efficient computation of coarse-grained velocity and scalar fields.
We apply the multi-scale framework to a large-eddy simulation of an idealised heterogeneous urban environment of 512 buildings based on a typical London height distribution. 
We observe that for this geometry, the characteristic urban lengthscale $\ell\approx 50$ m, which is the averaging lengthscale $L$ at which as much variance in the \review{mean} flow is resolved as is unresolved. 
For $L>400$ m, the statistics become approximately homogeneous, suggesting that non-building-resolving numerical weather prediction (NWP) models can be applied without modification at resolutions of 400 m and above for the case under consideration.
We derive the multi-scale planar- and Reynolds-averaged momentum equation and show that for neutral cases, NWP models require parameterisation of the distributed drag and the unresolved turbulence and dispersive stress.
An \review{\emph{a priori} analysis reveals that the} drag parameterisation from Sutzl \textit{et al.}, 2020, \textit{Bound-Layer Meteorol.}, \textbf{178}: 225–248 holds reasonably well for resolutions $L$ above 200 m. Below this value, the problem becomes inhomogeneous and the parameterisation works less well.
The unresolved stresses are well represented by a $k-\omega$ closure with a value of $\omega=0.4\, \U{s}^{-1}$.
However, an even more accurate closure can be derived from the Sutzl drag parameterisation that does not require further turbulence information.
\end{abstract}

\section{Introduction}
\label{intro}
The resolution of Numerical Weather Prediction (NWP) models has advanced significantly over recent decades, transitioning from synoptic-scale $O$(10 km) grids \citep{Bryan2003} to convection-permitting $O$(1 km) systems \citep{Baldauf2011, Tang2013}. This progress, driven by computational advancements, enables explicit resolution of critical atmospheric processes such as deep convection and orographic flows, reducing reliance on parameterisations \citep{Lean2024}. Kilometre-scale models are now operational for weather forecasting, regional climate downscaling \citep{Prein2015, Belusic2020}, and air quality assessments. However, a persistent limitation across NWP frameworks is the unresolved representation of urban morphology-buildings, vegetation, and surface heterogeneities that remain subgrid-scale features, necessitating parameterised approximations.

The urban canopy region is at the lowest level of the atmospheric boundary layer and is characterised by strong heterogeneity and complex interactions between the urban surface and the atmosphere. Traditional urban canopy models (UCMs) \review{aggregate} aerodynamic drag, radiative trapping, and energy exchange into bulk parameters such as roughness length, displacement height, and plan area density \review{that connect to the lowest atmospheric level directly \citep{Lipson2023}. Whilst this approach works reasonably well for areas with low-rise buildings, it becomes problematic for areas with tall buildings that extend substantially into the atmospheric boundary layer. Vertically distributed UCMs \citep{ Schoetter2020, Sutzl2021, Lu2024} improve momentum and scalar flux partitioning by distributing drag, heat over the vertical. However, both approaches assume horizontal homogeneity within grid cells, which breaks down below $O$(1 km) resolutions}, as urban landscapes exhibit heterogeneity at neighbourhood scales (about 100-500 m) due to variations in building height, street geometry, and land cover \citep{Stewart2012}. Such fine-scale variability drives localised microclimates, including urban heat islands, airflow channelling, and pollutant dispersion \citep{Masson2020}, which are poorly captured by current parameterisations.


High-resolution simulation solves some of these problems. Recent studies demonstrate that hectometric models (HMs) improve predictions of near-surface winds and turbulence kinetic energy in urban areas by up to 30\% compared to coarser models \citep{Lean2024}, highlighting their potential to reduce dependency on empirical parameterisations. Moreover, resolving neighbourhood-scale features partially enters the `turbulence grey zone', where subfilter turbulent fluxes coexist with resolved eddies \citep{Honnert2020}. At these scales, models can better represent terrain-driven phenomena (e.g., cold pools in valleys, slope winds) and urban surface-atmosphere interactions \citep{Brun2017, Smith2021}.

\review{However, at high horizontal NWP resolutions it will become necessary to relax the UCM assumption of horizontal homogeneity, and to understand how the vertical distribution depends on the horizontal resolution. In order to be able to provide the answer to this question, a multi-resolution framework is required, whereby high-resolution simulations that resolve individual buildings can be coarse-grained to arbitrary NWP resolution, such that it is possible to diagnose \emph{a priori} what parameterisation is needed to represent the buildings within the NWP model.
Coarse-graining is a well-established technique to study, e.g. grey-zone turbulence \citep{Honnert2020}, but there is no established multi-scale framework that can be used for the urban canopy.}


\review{Coarse-graining methods apply a spatial averaging, or filtering, operation on a fully resolved field in order to obtain a `smoothed' coarse-resolution field \citep{Wyngaard2004, Honnert2011}. 
Spatial averaging has been routinely used for urban canopies, inspired by the formalism presented for vegetation canopies \citep{Raupach1982, Finnigan2000b, Nepf2012}. 
Within the context of urban canopies, \citet{Schmid2019} provide a spatial-averaging framework based on the volume-averaging framework of \citet{Whitaker1999}, which was developed for porous media flows. A unique aspect of  NWP models is that near the surface (i.e.\ in the canopy layer), the horizontal resolution of NWP models is much larger than the vertical resolution, as variations in the vertical tend to be much stronger than in the horizontal. Indeed, typical NWP simulations have a resolution of $O$(1 m) in the vertical but $O$(1 km) in the horizontal; even in the context of a hectometric model, the cell aspect ratio will be at least 100. This suggests that an appropriate averaging region on the canopy layer can be defined as a thin slab, i.e.\ reducing a three-dimensional volume averaging to a two-dimensional planar averaging approach, which is the approach used in \citet{Schmid2019} and also for vegetation canopies \citep{Raupach1982}.}
\review{However, the \citet{Schmid2019} framework is not straightforward to apply directly to simulation data, since analyses are typically performed using averages over the entire domain and not for averages of arbitrary length scale $L$.
In addition, the approach is computationally inefficient and does not explicitly use a planar averaging region.}

The aim of this paper is to present a computationally efficient multi-resolution framework based on the \citet{Whitaker1999, Schmid2019} spatial averaging approach and to apply this framework to results from building-resolving large-eddy simulation.
This coarse-graining approach allows us to systematically study heterogeneity in the flow statistics.
In the context of NWP models, it facilitates analysis of the required drag and turbulence parameterisation for non-building resolving models at different horizontal resolutions. 
The paper is structured as follows. \S2 mainly introduces a multi-resolution spatial averaging framework to grain the coarse field from the original high-resolution one; \S3 introduces the simulation details of LES; and \S4 presents and discusses the multi-resolution results on drag force and turbulent momentum stress, respectively. The highlights are remarked in the \S5.

\section{Multi-resolution planar-averaging framework}
\label{sec:multiresolution}

\review{In this section, we derive a multi-resolution framework in terms of convolution filters.}
The main advantages of using convolution filters are: 1) that it standardises the methodology, bringing it close to the large-eddy simulation formalism \citep{pope_2000}; and 2) that convolutions can be computed efficiently using Fast-Fourier transforms (FFTs).
In addition, we formulate the spatial averaging purely in terms of a planar filter. The work in \citet{Schmid2019} is cast in planar form, but the integral representing the effect of the solid boundaries is still presented in the volume-averaging form. Here, we capitalise on recent work by \citet{Maarten_2021} who derive identities that permit a fully planar formulation. A description of how the volume averaging can be transformed to planar averaging and a derivation of the spatial averaging theorem using convolutions is provided in Appendix \ref{sec:volumeaveraging}. The main results needed for the convolution formalism are presented here. 

\subsection{Planar averaging using convolution filters}
\begin{figure}
    \centering
    \includegraphics[width = 14cm]{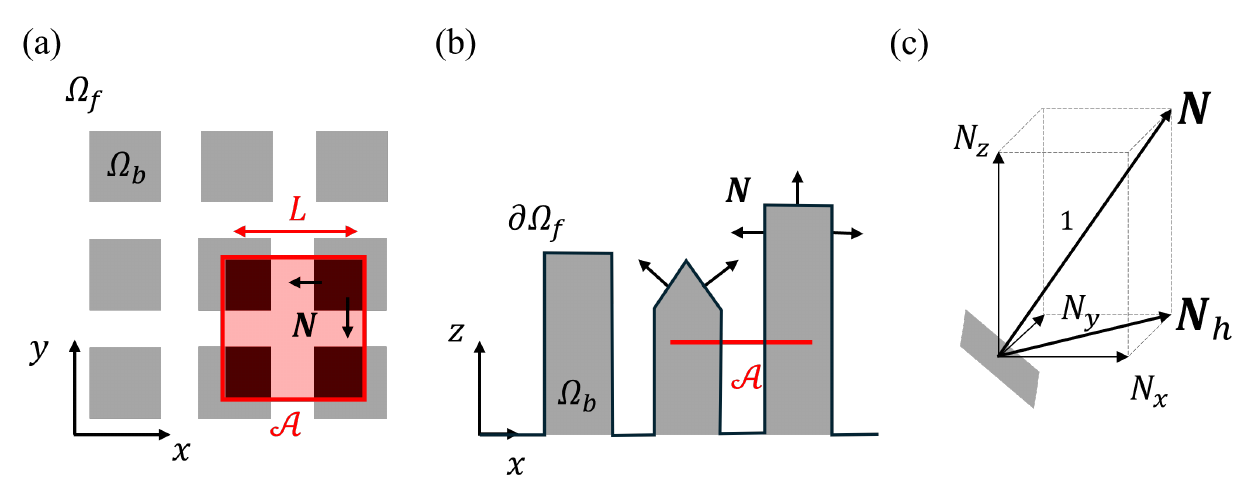}
    \caption{Definition sketch of domain and filter $\mathcal A$. (a) plan view, (b) elevation view. The domain of $\Omega$ is comprised of a fluid subdomain $\Omega_f$ (in white), a solid subdomain $\Omega_b$ (in grey), and a fluid-solid interface $\partial \Omega_f$ (solid black lines). The building-surface 3-D normal vectors $\vec N$ point into the fluid domain, and the decomposition is shown in (c). The 2-D square filter $\mathcal{A}$ with averaging length $L$ is shown in red.}
    \label{fig:sketch}
\end{figure}

The domain of interest $\Omega$ is comprised of a subdomain containing the fluid $\Omega_f$ and a solid subdomain $\Omega_b$ such that $\Omega = \Omega_f \cup \Omega_b$ (Fig. \ref{fig:sketch}). 
The boundary between the fluid and solid phase is denoted $\partial \Omega_f$, which has a (3-D) normal vector $\vec N$ pointing into the fluid domain. The horizontal, spanwise and vertical coordinates are denoted $x$, $y$ and $z$, respectively. 
Since averaging will take place over the streamwise directions, the coordinate vector $\vec x$ is denoted as $\vec x = [\vec x_\perp; z]$, where $\vec x_\perp=[x, y]^T$. 
Denoting the filter associated with the area-averaging $\mathcal A$, the superficial average \citep[also called comprehensive average, the average normalised by the total area including the solid phase; cf.][]{Xie2018} of an arbitrary scalar $\varphi$ is given by
\begin{equation}
 \label{eq:areaav}
 \sav{\varphi}(\vec x) \equiv  \int_{\Omega_f(\vec x)} \mathcal A(\vec x_\perp - \vec y_\perp) \varphi(\vec y_\perp, z) \d \vec y_\perp \, .
\end{equation}

The presence of $\vec x$ in the integration domain $\Omega_f(\vec x)$ means that the integration region comprises all the fluid-occupied areas within the support of the filter. 
This can be made explicit by defining a mask function $I_f(\vec x)$ which is 1 when $\vec x \in \Omega_f$ and 0 otherwise, which allows the convolution integral above to be expressed as 
\begin{equation} \label{eq: convolution}
 \volav{\varphi}(\vec x) =  \int_{\Omega(z)} \mathcal A(\vec x_\perp - \vec y_\perp)  I_f(\vec y_\perp, z) \varphi(\vec y_\perp, z) \d \vec y_\perp \, .
\end{equation}
This is a standard planar convolution integral over the entire domain $\Omega$ at given $z$ which can be evaluated efficiently using the convolution theorem and Fast-Fourier transforms \citep{Bracewell2000}.

Formally, there are few restrictions on the filter, except that it satisfies the normalisation condition $\int \mathcal A(\vec x_\perp) \d \vec x_\perp = 1$. Here, a symmetric square filter is used, defined as
\begin{equation}
  \label{eq:Arect}
  \mathcal A(x,y) \equiv 
  \begin{cases}
      L^{-2} & \textnormal{if } \left(-\frac{L}{2} < x < \frac{L}{2} \right) \cup \left( -\frac{L}{2} < y < \frac{L}{2} \right) \, ,\\
      0 & \textnormal{otherwise} \, ,
  \end{cases}
\end{equation}
where $L$ is the averaging length scale. 
Note that $\sav{\sav X} \ne \sav X$ in general. However, if the underlying characteristic length scale of the urban surface $\ell$ is much smaller than the averaging length scale $L$, that is $L\gg \ell$, we expect  $\sav{\sav X} \approx \sav X$. In other words, $\sav{\sav X} \approx \sav X$ if there is a separation of scales between the characteristic urban lengthscale $\ell$ and the averaging lengthscale $L$ \citep{Schmid2019}. 

In the filter formalism, the area  (planar) fraction $\varepsilon=A_f/L^2$, where $A_f$ is the fluid surface within the averaging region, is defined as
\begin{equation}
\varepsilon(\vec x) \equiv \int_{\Omega_f(\vec x)} \mathcal A(\vec x_\perp - \vec y_\perp) \d \vec y_\perp.
\end{equation}
This equation shows the interaction between the filter $\mathcal A$ and the integration over the fluid domain $\Omega_f$, which may only cover part of the area over which $\mathcal A$ filters.
The porosity $\varepsilon$ can be used to calculate the intrinsic average $\iav{\varphi}$ (the average normalised by only the area of the variable fluid phase), indicative of the average value of $\varphi$ inside the fluid phase:
\begin{equation}
 \iav{\varphi} = \varepsilon^{-1} \sav{\varphi}.
\end{equation}

\subsection{Differentiation rules}

Since the solid phase is stationary, the spatial averages commute with time differentiation.
The same is not true for spatial differentiation. 
For the planar filter $\mathcal A$, the scalar spatial averaging theorem \citep[][ eq.\ 1.2-15]{Whitaker1999} is given by (see appendix \ref{sec:volumeaveraging}):
\begin{equation} \label{eq:spdiff}
  \sav{\frac{\partial \varphi}{\partial x_i}}
  = \frac{\partial \sav{\varphi}}{\partial x_i} 
  - \oint_{\partial \Omega_f} \mathcal A \varphi \frac{N_i}{|\vec N_\perp|} \, \d s \, .
\end{equation}
Here, the last term is a \review{Leibniz-like} horizontal line integral over the fluid-solid boundary $\partial \Omega_f$ at given $z$ within the support of $\mathcal A$. 
The negative sign of the line integral is due to the (3D) surface normals $\vec N$ pointing into the domain \review{(the direction of the 3D vector $\vec N$ is shown by the arrow in Fig. \ref{fig:sketch}c)}. 
The quantity $\vec N_\perp = [N_x, N_y]^T$ is the component of $\vec N$ in the horizontal $x-y$ plane. 
The term $|\vec N_\perp|$ accounts for the local surface orientation that need not be aligned with the horizontal direction. 
\review{For vertical surfaces, $|\vec N_\perp| = 1$ (e.g., the horizontal vectors on the last building in Fig. \ref{fig:sketch}b), for angled surfaces $0<|\vec N_\perp|<1$ (e.g., the vectors on the middle building in Fig. \ref{fig:sketch}b), and for horizontal surfaces, for example, flat roofs, $|\vec N_\perp|=0$ (the vertical vector on the last building in Fig. \ref{fig:sketch}b).} 
Superficially, it seems the integral is undefined for horizontal surfaces, but it can be shown \citep{Maarten_2021} that the integral becomes finite upon integrating over $z$. Physically, one can interpret this as a flat surface providing a finite contribution over an infinitesimal distance in $z$. 
Note that the presence of these apparent singularities is an unavoidable consequence of using a planar average rather than a volumetric average, \review{and its evaluation is discussed in \S \ref{sec:implementation}.
To demonstrate the physical meaning of the line integral, it is useful to apply \eqref{eq:spdiff} to the mean pressure $\overline p$. The horizontal pressure gradient then becomes 
\begin{equation*} 
  \sav{\frac{\partial \overline p}{\partial x}}
  = \frac{\partial \sav{\overline p}}{\partial x} 
  - \oint_{\partial \Omega_f} \mathcal A \overline p \frac{N_x}{|\vec N_\perp|} \, \d s \, .
\end{equation*}
This demonstrates that the filtering operation produces the form drag in the $x$-direction within the filter support. Note that under the assumption of flow in the positive $x$-direction and cuboids that face the wind head-on, we would have $N_x / | \vec{N}_\perp|=-1$ on windward vertical surfaces, +1 on leeward vertical surfaces and 0 on sidewalls. The top surface of buildings will not contribute to the form drag since $N_x = 0$, although care must be taken to show that the apparent singularity caused by $| \vec{N}_\perp |$ = 0 does not cause trouble (see \S \ref{sec:implementation}).
}

For the planar filter $\mathcal A$, the vector form of the spatial averaging theorem \citep[][ eq.\ 1.2-23]{Whitaker1999} is given by (see appendix \ref{sec:volumeaveraging})
\begin{equation} \label{eq:spdiff_vec}
  \sav{\frac{\partial F_i}{\partial x_i}}
  = \frac{\partial \sav{F_i}}{\partial x_i}
  - \oint_{\partial \Omega_f} \mathcal A \frac{F_i N_i}{|\vec N_\perp|} \d s \, ,
\end{equation}
where repeated indices imply implicit summation.
\review{As an example of the application of this theorem, consider filtering the mean continuity equation $\nabla \cdot \overline{\vec u} = 0$. Using \eqref{eq:spdiff_vec} with $F_i = \overline u_i$ results in
\begin{equation*}   \sav{\frac{\partial \overline u_i}{\partial x_i}}
  = \frac{\partial \sav{\overline u_i}}{\partial x_i}
  - \oint_{\partial \Omega_f} \mathcal A \frac{\overline u_i N_i}{|\vec N_\perp|} \d s   = \frac{\partial \sav{\overline u_i}}{\partial x_i}
\, ,
\end{equation*}
since $\overline u_i N_i = 0$ on $\partial \Omega_f$ due to the impermeability of the buildings. Therefore, $\nabla \cdot \sav{\overline{\vec u}} = 0$, implying that the filtered velocity field is also divergence-free.
}

If the averaging length $L$ is much larger than the characteristic length scale $\ell$ of the urban surface, the average concentration $\sav \varphi$ and horizontal fluxes $\sav{F_x}$ and $\sav{F_y}$ will be independent of $x,y$, implying that the spatial averaging theorems reduce to \citep{Schmid2019}
\begin{align}
\sav{\frac{\partial \varphi}{\partial x_i}}
  &= \frac{\partial \sav \varphi}{\partial z}
  - \oint_{\partial \Omega_f} \mathcal A \varphi \frac{N_i}{|\vec N_\perp|} \, \d s \, ,  \label{eq: idt1}\\
\sav{\frac{\partial F_i}{\partial x_i}}
  &= \frac{\partial \sav{F_z}}{\partial z}
  - \oint_{\partial \Omega_f} \mathcal A \frac{F_i N_i}{|\vec N_\perp|} \, \d s \label{eq: idt2}.
\end{align}

\subsection{Triple decomposition}

We define the Reynolds average of $\varphi$ as an ensemble average \citep{Schmid2019}: 
\begin{equation}
\label{eq:timeav}
    \overline{\varphi}(\vec x,t) \equiv \sum_{i=1}^{N} \varphi^{(i)}(\vec x,t) \, .
\end{equation}
Deviations from the average value $\overline \varphi$ are denoted $\varphi^\prime(\vec x, t) = \varphi(\vec x, t) - \overline \varphi(\vec x, t)$.
For statistically steady problems, time-averaging can be used in lieu of an ensemble average.
In the current paper, we will assume that Reynolds-averaging has been performed before spatial averaging, which avoids some of the complexities of simultaneous Reynolds and spatial averaging \citep{Schmid2019}.
We thus decompose the average velocity into the local spatial average and its deviation from the average as $\overline \varphi = \iav {\overline \varphi} + \overline \varphi^{\prime\prime}$. 
This decomposition is only defined in the fluid domain, and the intrinsic average is used since this represents the average value inside the fluid domain.
Substitution into Eq. \eqref{eq:timeav} results in the triple decomposition
\begin{equation} 
\label{eq:triplede}
    \varphi(\vec x, t) = \underbrace{\iav{\overline{\varphi}}(\vec x, t) + \overline{\varphi}^{\prime \prime}(\vec x, t)}_{\overline{\varphi}(\vec x, t)} + \varphi^{\prime}(\vec x, t) \, .
\end{equation}
For products, e.g.\ of a scalar flux  $u_i \varphi$, this implies
\begin{equation} \label{eq: decomp_wphi}
\begin{split}
  \sav{\overline{u_i\varphi}} 
  &= \sav{\overline u_i\, \overline \varphi} + \sav{\overline{u_i^\prime \varphi^\prime}} \\
  &= \sav{\iav{\overline{u_i}} \iav{\overline{\varphi}}} +
  \sav{\iav{\overline{u_i}} \overline \varphi^{\prime\prime}} +
  \sav{\overline u_i^{\prime\prime} \iav{\overline{\varphi}}} +
  \sav{\overline u_i^{\prime\prime} \overline \varphi^{\prime\prime}} +
  \sav{\overline{u_i^\prime \varphi^\prime}} \, .\\
\end{split}
\end{equation}
%

\subsection{Reynolds- and area-averaged streamwise momentum equations} \label{sec: Re-u-momentum equation}
Ignoring buoyancy effects and assuming incompressible airflow, the (unsteady) Reynolds-averaged streamwise horizontal momentum equation is given by 
\begin{equation}  \label{eq:URANS}
  \frac{\partial \overline u}{\partial t} +\frac{\partial \overline u_j \overline u}{\partial x_j} +  \frac{\partial \overline{u_j' u'}}{\partial x_j}
  +\frac{\partial \overline p}{\partial x} 
  = f+\nu \frac{\partial^2 \overline{u}}{\partial x_j^2}  \, ,
\end{equation}
where $\overline{u}$ represents the mean velocity in the streamwise $x$-direction, $\overline{p}$ is the kinematic deviatoric pressure, and $f$ is the volumetric forcing term, typically due to a pressure gradient.
This equation can be written in flux form as
\begin{equation}
  \frac{\partial \overline u}{\partial t} +\frac{\partial F_i}{\partial x_i} = f, \quad \textnormal{where }  \quad
  F_i = \overline{u}_i \overline u + \overline{u^\prime_i u^\prime} - \nu \frac{\partial \overline u}{\partial x_i} + \overline{p} \delta_{i1}
\end{equation}
where $\delta_{ij}$ is the Kronecker delta.
Superficial averaging and application of the spatial averaging theorem Eq. \eqref{eq:spdiff_vec} results in 
\begin{equation} \label{volav_s}
  \frac{\partial \sav{\overline u}}{\partial t} +
\frac{\partial \sav{F_i}}{\partial x_i}
   = \sav{f} - \sav{f_D},
\end{equation}
where $\sav f = \varepsilon f$ and the drag force induced by the buildings $\sav{ f_D}$ is given by 
\begin{equation} \label{eq:slab_drag}
    \sav{f_{D}} = 
    -\oint_{\partial \Omega_f} \mathcal A \frac{F_i N_i}{|\vec N_\perp|} \d s = 
    -\oint_{\partial \Omega_f} \mathcal A \, \left( \overline p \frac{  N_x}{|\vec N_\perp|} - \nu \frac{\partial \overline u}{\partial x_j}  \frac{ N_j}{|\vec N_\perp|} \right)\d s \, .
\end{equation}
\review{This equation explicitly provides the drag force induced by the building surfaces that is comprised of a form drag associated with the pressure and a skin drag associated with the viscosity \citep{Raupach1982, Finnigan2000b, Nepf2012, Sutzl2020}. 
}

Considering the fact that NWP models take into account the net effect of buildings without actually resolving them, the governing equations are posed in terms of \emph{superficial} quantities. 
Recasting Eq. \eqref{volav_s} in the standard form for the streamwise momentum equation, we thus obtain:
\begin{equation} \label{eq:NWP}
  \frac{\partial \sav{\overline u}}{\partial t} +
  \frac{\partial \sav{\overline{u}_i}\sav{\overline u}}
  {\partial x_i}
  + \frac{\partial \sav{\overline p}}{\partial x} 
   = \sav{f} - \sav{f_D} 
+ \frac{\partial \sav{\tau_i}}{\partial x_i},
\end{equation}
where 
\begin{equation} \label{eq: tau_i}
\sav{\tau_i} = \sav{\overline{u}_i}\sav{\overline u}- 
\sav{\overline{u}_i \overline u} 
- \sav{\overline{u^\prime_i u^\prime}} \, ,
\end{equation}
is the unresolved kinematic stress term. Note that we have not decomposed the term $\sav{\overline{u}_i \overline u}$ into its components and that the viscous flux has been omitted since it is negligible in magnitude compared to the other fluxes.
Equation \eqref{eq:NWP} highlights the two terms that need to be closed in order to represent the effect of buildings in a non-building resolving NWP model: 1) the distributed drag contribution $f_D$; and 2) the unresolved kinematic stress $\tau_i$.

\subsection{Statistically steady flow with periodic boundary conditions} \label{sec: steady_flow}
The case considered in this paper is a turbulent urban boundary layer flow with periodic boundary conditions and a free-slip top boundary at $z=h$ in a statistically steady state. In this situation, Eq. \eqref{eq:NWP} becomes
\begin{equation} 
  \frac{\partial \sav{\overline{u}_i}\sav{\overline u}}
  {\partial x_i}
  + \frac{\partial \sav{\overline p}}{\partial x} 
   = \sav{f} - \sav{f_D}
+ \frac{\partial \sav{\tau_i}}{\partial x_i},
\end{equation}
which is not much of a simplification. No term except the time derivative can be assumed zero for an arbitrary averaging length $L$. 
However, if the averaging length scale is very large, the problem will become homogeneous in $x$ and $y$ and all the horizontal derivatives will vanish. With this in mind, it is useful to rewrite the equation above as 
\begin{equation} 
  \label{eq:heterogeneityeffects}
  \frac{\partial \sav{\overline{u}}\sav{\overline u}}
  {\partial x}+  \frac{\partial \sav{\overline{v}}\sav{\overline u}}
  {\partial y}+ \frac{\partial \sav{\overline p}}{\partial x} - \frac{\partial \sav{\tau_x}}{\partial x}
 - \frac{\partial \sav{\tau_y}}{\partial y}
   = \sav{f} - \sav{f_D}- \frac{\partial \sav{\overline{w}}\sav{\overline u}}
  {\partial z} + \frac{\partial \sav{\tau_z}}{\partial z} \, ,
\end{equation}
Here, the horizontal terms on the left-hand side will vanish if the problem is homogeneous, leaving the terms on the right-hand side of the equation to balance.
The equation above indicates that an imbalance in the terms on the right-hand side implies inhomogeneity in $x$ and $y$ via horizontal terms on the left-hand side and vice versa.  

Homogeneity is expected when $L\gg \ell$, i.e. $L$ is much larger than the characteristic urban lengthscale $\ell$. In this case, the horizontal momentum equation simplifies to 
\begin{equation} \label{eq:NWP2}
   \frac{\d \sav{\tau_z}_\infty}{\d z} =  \sav{f_D}_\infty - \sav{f}_\infty\, ,
\end{equation}
where
\begin{equation}
\sav{\tau_z}_\infty =  - 
\sav{\overline{w} \, \overline u}_\infty 
- \sav{\overline{w^\prime u^\prime}}_\infty  = - \sav{\overline w^{\prime\prime} \overline u^{\prime\prime}}_\infty
-\sav{\overline{w^\prime u^\prime}}_\infty \, ,
\end{equation}
note that $\sav{\overline{w}}_{\infty} = 0$ due to volume conservation. The subscript on the superficial average indicates the averaging length $L$, in this case $L/\ell \gg 1$ which we denote as $L= \infty$. In the limit of $L\gg\ell$, the quantity $\sav{\varphi}_\infty$
is identical to the conventional plane-average $\spav \varphi$, which is typically used in computational studies \citep{Coceal2006,Xie2008,Giometto2016, Nazarian2020, Sutzl2020}. The superficial plane average of $\varphi$ is defined as
\begin{equation} \label{eq:slabav}
    \spav{\varphi}(z, t) \equiv \frac{1}{A_T} \int_{\Omega_f(z)} \varphi(\vec x, t) \, \d \vec x_\perp \, ,
\end{equation}
where $A_T = L_x L_y$ is the total area of the horizontal plane. 
Note that the plane average is distributive and that contrary to $\sav{\varphi}$, $\spav{\spav{\varphi}} = \spav{\varphi}$, i.e., this operator behaves like a regular Reynolds average.

Integrating Eq. \eqref{eq:NWP2} from height $z$ to the domain top at height $h$ results in the following cumulative stress relation \citep{Sutzl2020}:
\begin{equation} \label{eq: int_force_balance}
  \sav{\tau_z}_\infty(z) = \tau_{f;\infty}(z) - \tau_{D;\infty}(z) \, ,
\end{equation}
where
\begin{align*}
    \tau_{f;L}(\vec x_\perp, z) & \equiv \int_z^h \sav{f}_L(\vec x_\perp, z') \d z' \, , \\ 
    \tau_{D;L}(\vec x_\perp, z) & \equiv \int_z^h \sav{f_{D}}_L(\vec x_\perp, z') \d z' \, ,     
\end{align*}
are the cumulative shear stresses due to forcing and drag, respectively. They have been formulated here for an arbitrary length scale $L$. For $L\gg \ell$, the stresses are a function of $z$ only. 
At the ground surface $z = 0$, the relation reduces to $\tau_{f;\infty}(0) = \tau_{D;\infty}(0) \equiv \tau_0$, where ${\tau_0}$ is the kinematic surface shear stress \citep{pope_2000}. 

In order to keep consistency in notation with earlier work, we will use that in the absence of a specific filter size in the subscript of a quantity, it is implied that $L=\infty$, e.g. $\tau_f \equiv \tau_{f;\infty}$; and we will use the conventional plane average notation $\spav{\varphi}$ instead of its equivalent $\sav{\varphi}_{\infty}$. With this in mind, the cumulative stress relation Eq. \eqref{eq: int_force_balance} is simply 
\begin{equation}
   \label{eq: int_force_balance2}
    \spav{\tau_z} = \tau_{f}(z) - \tau_{D}(z).
\end{equation}

\section{Simulation details}
\begin{figure}
    \centering
    \includegraphics[width=14cm]{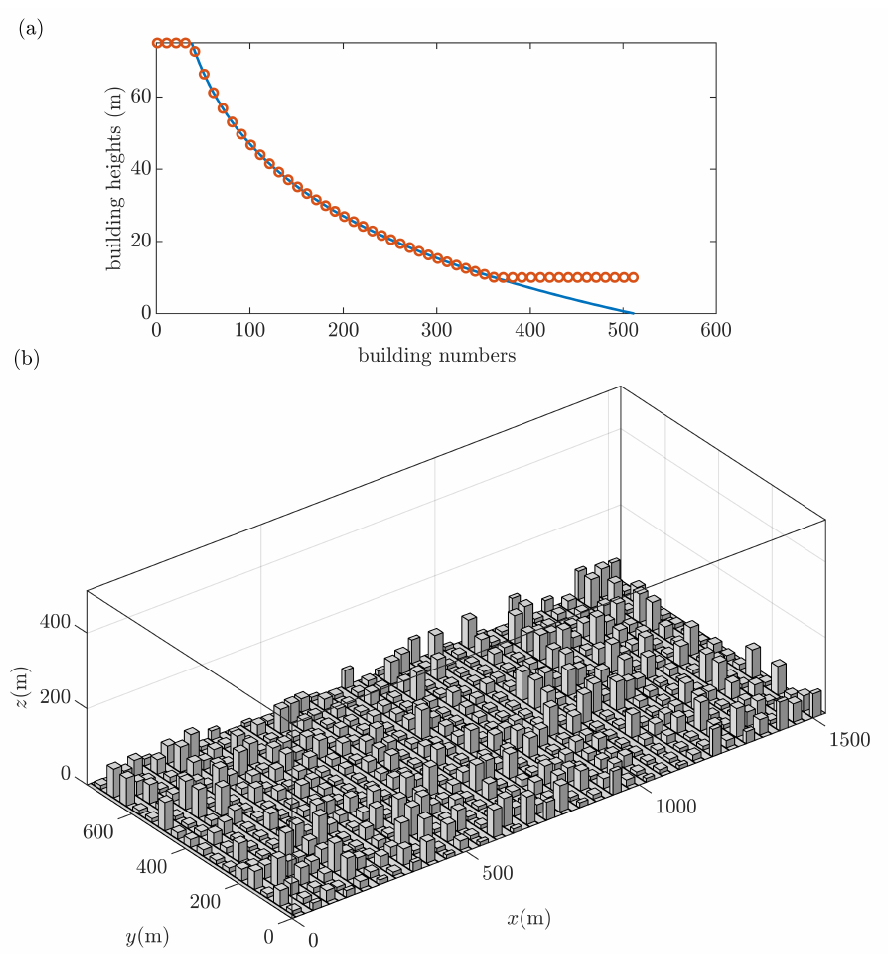}
    \caption{Construction of the urban geometry. (a) Determination of building heights from $L_b(z)/W$ (solid line) by sampling $L_b/W$ at integer values (circles; every ten buildings shown); (b) The synthetic urban geometry.}
    \label{fig:morphology}
\end{figure}

\subsection{Case setup} \label{sec: case setup}
In order to investigate multi-scale aspects of urban flows, we consider an idealised quasi-realistic urban morphology that has a staggered arrangement of cuboidal buildings of breadth and width $W=24 \, \U m$, whose height $H$ is random (Fig. \ref{fig:morphology}).
This case is inspired by  \citet{Xie2008}, but with a much larger simulation domain of $L_x \times L_y \times L_z = 1\,536\,\U{m} \times 768\,\U{m} \times 512\,\U{m}$, and using a height distribution representative for London \citep{Sutzl2021} rather than a Gaussian height distribution.

With a spacing of size $W$ between the buildings, the domain will contain $N_b = 32 \times 16 = 512$ buildings. This implies that the plan-area index $\lambda_p \equiv A_p / A_T = 0.25$ where the plan area and total area are defined as, respectively, $A_p =N_b W^2$ and $A_T = L_x L_y$.
For the distribution of building heights $H$, we use an empirical relation which was obtained using metre-scale GIS data for the Greater London area \citep{Sutzl2021}.
It was found that the total building width $L_b$ (i.e.\ the sum of the frontal widths of all buildings) at elevation $z$ can be approximated as:
\begin{equation} \label{eq:total_width}
    L_b(z; \alpha) =  \frac{A_F}{ h_{\max}}\frac{\alpha e^{-\alpha z/h_{\max}}}{1-e^{-\alpha}} \, , \quad \quad z \leq h_{\max}\, ,
\end{equation}
where the coefficient $\alpha(r)$ was observed to depend on the maximum-to-mean-building height  $r = h_{\max}/h_m$ as
\begin{equation}
    \alpha(r) = 1.355 r - 0.7807 \, .
\end{equation}
Here, $h_{\max}$ is the maximum building height, $h_m$ is the mean building height, and $A_F$ is the total frontal area of the buildings. Note that $A_F/h_{\max}$ is the total building width $L_b$ if all buildings would have height $h_{\max}$. 
Note that above the maximum building height $L_b = 0$.

The mean building height is set to $h_m = 30\, \U m$ and the ratio is set to  $r= 2.5$, both of which are close to the average value of London reported by \citet{Sutzl2021}.
Thus the maximum building height is $h_{\max} = 75\,\U m$ and $\alpha = 2.61$. 
At $z=0$, the total building width $L_b(z; \alpha)=N_b W$, which implies that the frontal area can be determined from Eq. \eqref{eq:total_width} as
\begin{equation}
    A_F =  \frac{1-e^{-\alpha}}{\alpha} N_b W h_{\max}, 
\end{equation}
 resulting in $A_F = 327 \, 138 \,\U{m}^2$ and thus a frontal area index 
 $\lambda_f \equiv A_F/A_T = 0.28$.

In order to assign heights to each cuboidal building, note that Eq. \eqref{eq:total_width} divided by $W$ can be inverted to show the height $z$ as a function of the number of buildings $L_b(z)/W$, which represents the distribution of heights as a function of the number of buildings (Fig. \ref{fig:morphology}(a), solid line). 
The building heights can be determined by uniformly sampling this function for each $i =1,2,\ldots, N_b$. The first X buildings, where the function is not defined, are set to $h_{\max}$, and the minimum building height is set to $10$ m. As a final step, the heights are randomly distributed using a uniform distribution to all $512$ buildings as shown in the morphology in Fig. \ref{fig:morphology}(b). The total building width $L_b$ is shown in Fig. \ref{fig:zeta-tau relation}(a) against the height, above $z = 10$ m which is the minimum building height, the building width $L_b$, also indicating that the distributive frontal area smoothly decreases to zero at the top of the canopy region. Due to the minimum building height setting, the actual frontal area index is $\lambda_f = 0.29$, slightly greater than previously designed.

The urban flows over this morphology are performed by the large-eddy simulation using our open-source code uDALES \citep{Owens2024}, presenting the solid boundary using the immersed boundary method (IBM). The dynamics near the boundary are parametrised by the logarithmic wall functions \citep{Uno1995, Suter2022}. The turbulent eddy viscosity is calculated following the \citet{Vreman2004} subgrid model. The code employs a second-order central difference scheme on a staggered Arakawa C-grid for spatial discretisation and an explicit third-order Runge-Kutta scheme for time integration.

The simulation is neutral and a constant pressure gradient forcing $\d P/\d x= 0.0042$ kg m$^{-2}$s${^{-2}}$ is imposed on the streamwise ($x$) direction to drive the wind. The simulation has a grid number of $N_x \times N_y \times N_z = 1\,024 \times 512 \times 512$ (i.e., the grid size is $\Delta x \times \Delta y \times \Delta z = 1.5 \,\U m \times 1.5 \,\U m \times 1 \,\U m$), with periodic boundary conditions applied in the lateral sides while the domain top is free-slip. The simulation runs for 10\,000 s and the last 8\,000 s is used to obtain converged time-averaged statistics.

\subsection{Implementation details}
\label{sec:implementation}
Efficient computation of the convolution integral Eq. \eqref{eq: convolution} is crucial to carry out a multi-scale analysis. In the spatial domain, the convolution integral needs to be evaluated $N_x \times N_y \times N_z$ times for a single averaging length $L$ which is computationally very expensive, particularly for high-resolution simulations: \review{assuming a filter contains $n^2$ grid cells, the convolution over the entire domain requires the number of operations of $O(n^2N_x^2N_z)$ where we have assumed $N_x = N_y$. This is acceptable for small filter scales but becomes computationally inhibiting for large filter sizes.} 
Therefore, we perform the convolution in the Fourier space by making use of the convolution theorem, which states that a convolution becomes a multiplication in Fourier space \citep{Bracewell2000}, \review{thus reducing the computational complexity of a 1-D convolution to $O(N_x \log N_x)$.} 
We thus compute the horizontal 2-D FFT of $I_f\varphi$ and the filter $\mathcal{A}$, multiply them, and then transform the product back to physical space using an inverse FFT, obtaining the filtered results for all points in the plane simultaneously.
\review{
Repeating this for all $N_z$ planes requires $O((\log N_x)^2 N_x^2N_z)$ operations, thus greatly reducing the computational cost, particularly for large simulation domains and large filters.
The algorithm can be downloaded from \citet{vanReeuwijk_Huang_2025}.
}
The filtering is performed for nine different averaging lengths: $L = 3, 6, 12, 24, 48, 96, 192, 384,$ and $ 768$ m.

Calculating the drag force Eq. \eqref{eq:slab_drag} also requires care as the line integral to evaluate since the building surfaces contain many horizontal surfaces for which $|\vec N_\perp|=0$. In order to ensure an appropriate discretisation of this term, note that by definition, for any boundary quantity $\phi$,
\begin{equation}
  \int \oint_{\partial \Omega_f(z)} \frac{\phi}{|\vec N_\perp|} \d s \d z = \oiint_{\partial \Omega_f} \phi \d S,
\end{equation}
i.e. integrating the line integral over $z$ results in the total flux exchanged across the solid-fluid surface. In a discrete form, this equation is given by
\begin{equation}
\sum_k\smash{ \left( \sum_{m:K_m=k} \frac{\phi_m}{|\vec N_\perp|_m} \Delta s_m \right)} \Delta z_k = \sum_m  \phi_m A_m,
\end{equation}
where each cell-facet $m$ has area $A_m$, surface normal $\vec N_m$, and its contribution $\phi_m$ is linked to a grid cell at index $I_m$, $J_m$ and $K_m$. Since both sides of this equation use all cell facets exactly once, it follows that
\begin{equation}
  \frac{\Delta s_m}{|\vec N_\perp|_m} = \frac{A_m}{\Delta z_{K_m}},
\end{equation}
which clarifies how ${\Delta s_m}/{|\vec N_\perp|_m} $ needs to be evaluated; it also shows that this term is always nonzero and finite. 
In order to be able to straightforwardly incorporate filtering, the surface quantity $\phi$ is converted to a surface density field $\rho_\phi$ that contains the boundary term and which is only non-zero in the cells next to the boundary:
\begin{equation}
  \label{eq:Phi}
  \rho_{\phi;ijk} = \sum_{m \in M_{ijk}} \frac{\phi_m A_m}{\Delta x \Delta y \Delta z_{K_m}}
\end{equation}
where $M_{ijk} \in \{m: I_m = i, J_m = j, K_m = k \}$ is the set of all cell-facets associated with cell $i,j,k$.
Using Eq. \eqref{eq:Phi}, any surface quantity can be filtered straightforwardly using $\sav{\rho_\phi}(\vec x) = \int_{\Omega_f} \mathcal A(\vec x_\perp - \vec y_\perp) \rho_\phi(\vec y_\perp) \d \vec y_\perp$. Here we introduce a volumetric field for two surface quantities:
\begin{itemize}
    \item the distributed drag term, by setting $\phi_m = -(\overline p_m \vec e_x  - \nu (\nabla \overline u)_m) \cdot \vec N_m$. Denoting the volumetric drag density by $\rho_D$, we have that $\sav{f_D} = \sav{\rho_D}$.
    Here it should be noted that $\nu (\nabla \overline u)_m$ follows from the wall shear stress as calculated by the wall functions and not the resolved velocity field \citep{Owens2024}.  
\item the frontal area term, by setting $\phi_m = - \min(\vec e_u \cdot \vec N_m, 0)$, where $\vec e_u$ is a unit vector for the wind direction (where $e_{u;z}=0$).
This term can be understood by realising that the projected area of facet area $m$ is given as $A_m |\vec e_u \cdot \vec N_m|$ and that a contribution is only counted when it involves the windward side, i.e. $\vec e_u \cdot \vec N_m < 0$. 
Denoting the resulting frontal-area density $\rho_L$, we have that $\int \rho_L \d x \d y = L_b$ and $\int \rho_L \d V = A_F$.
\end{itemize}
\section{Results} \label{sec:1}
\subsection{\review{Classical plane-averaged} analysis} \label{sec: Classical analysis}
\begin{figure}
    \centering
    \includegraphics[width=15cm]{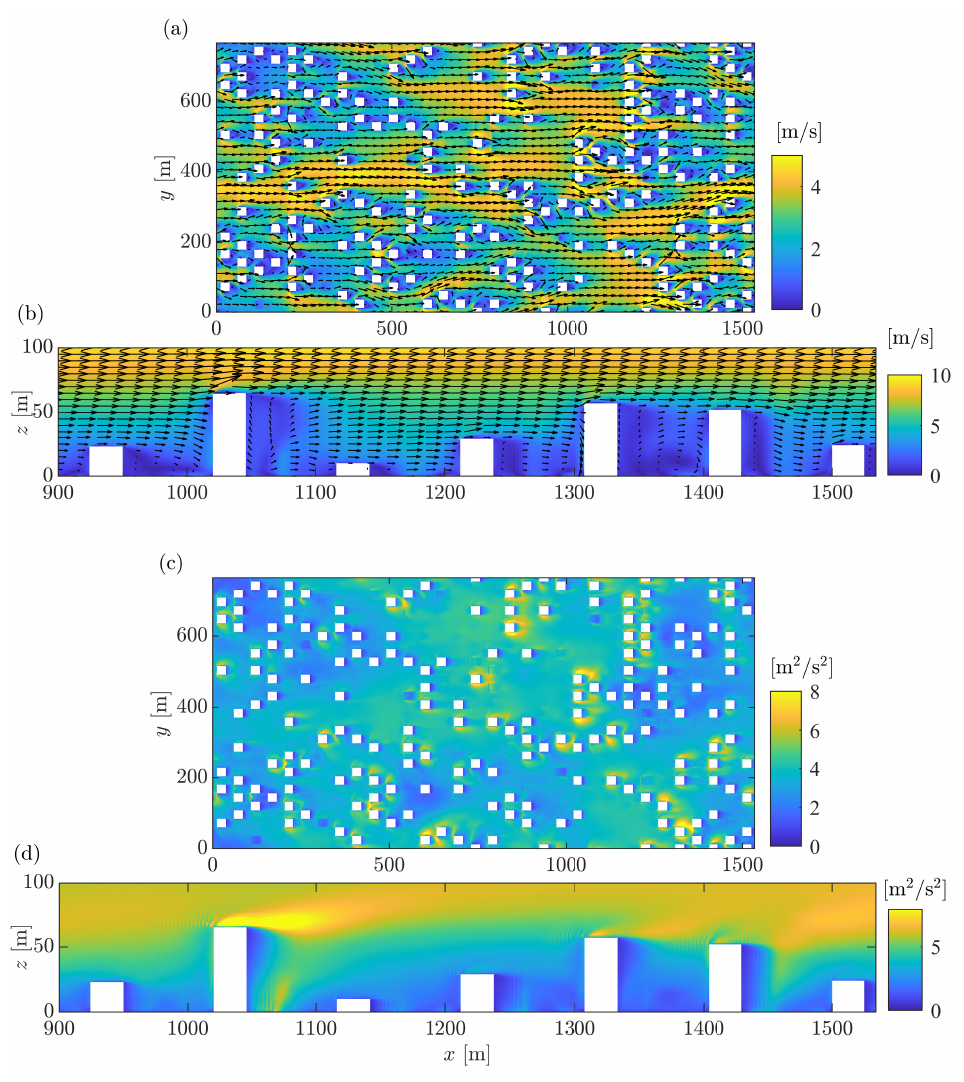}
    \caption{Time-averaged flow statistics. (a) Wind speed $\sqrt{\overline{u}^2 + \overline{v}^2}$ at mean building height $z = 30$ m overlaid with  $(\overline{u}, \overline{v})$ velocity vectors; (b) wind speed $\sqrt{\overline{u}^2 + \overline{w}^2}$  at the central vertical plane $y = 384$ m overlaid with  ($\overline{u}$,$\overline{w}$) velocity vectors;  (c) Turbulent kinetic energy $k$ at mean building height $z = 30$ m. (d) Turbulent kinetic energy $k$ at the central vertical plane $y = 384$ m. The white boxes represent the buildings. }
    \label{fig:velocity field}
\end{figure}

\begin{figure}
    \centering
    \includegraphics[width=14cm]{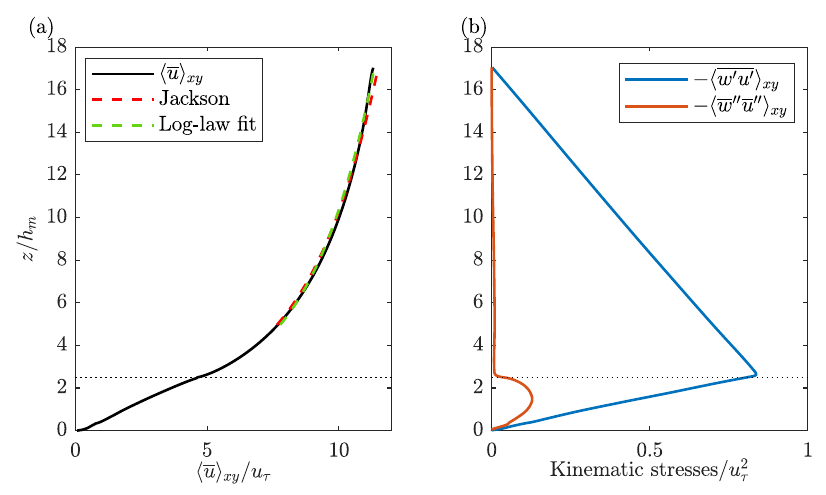}
    \caption{Plane-averaged flow statistics. (a) streamwise velocity $\spav{\overline{u}}$ against normalised height $z/h_m$, overlaid with logarithmic profiles Eq. \eqref{eq:loglaw} estimated using the log-law fitting and the Jackson displacement method; (b) normalised kinematic turbulent shear stress $-\spav{\overline{w^\prime u^\prime}}$ and dispersive stress $-\spav{\overline{w}^{\prime \prime} \overline{u}^{\prime \prime}}$. The dotted line marks the maximum building height $h_{\max}$.}
    \label{fig:slab-averaged u}
\end{figure}

The time-averaged wind speed at the mean building height $z=30$ m (Fig. \ref{fig:velocity field}a) shows large variations in wind speed and channelling in regions where buildings are absent.
A strong reduction in wind speed is shown near some of the tall building clusters \citep{Mishra2023, Mishra2024}, with wakes clearly visible downstream of the tall buildings.
The elevation view (Fig. \ref{fig:velocity field}b) shows the general increase of wind speed with height and also demonstrates the homogenisation of the wind in the $x$ direction with increasing height, consistent with the extent of the roughness sublayer (the height of this layer is $h_{\max}$ as will be shown in Fig. \ref{fig:slab-averaged u}). The velocity vectors show a clearly visible wake behind the building (in particular behind the first building in Fig. \ref{fig:velocity field}b). Accelerations are visible near the upwind top of the building (e.g., for the second cube in Fig. \ref{fig:velocity field}b), which is consistent with \citet{Coceal2007}. \review{In Fig. \ref{fig:velocity field}(b), the skimming flow can be observed in places, e.g., near the last cube; and also the flow recirculation, e.g., between the last third and last second cube.}

Figures \ref{fig:velocity field}(c, d) show the plane view at $z=30$ m and the vertical view at the central plane of time-averaged turbulent kinetic energy (TKE, $k(\vec x) \equiv \frac{1}{2} \left(\overline{u'u'}+\overline{v'v'}+\overline{w'w'} \right)$), respectively. \review{Note that the TKE field is correlated with the mean wind speed field: in the building cluster area, e.g. at the top right of Fig. \ref{fig:velocity field}(c), it can be seen to have low levels of turbulence due to the low velocity in the building wake; while at the upwind top of the building (e.g., the second building in Fig. \ref{fig:velocity field}d), large TKE can be observed due to the acceleration and large shear at the top.}


\review{Figure \ref{fig:slab-averaged u} presents the streamwise velocity and turbulent transport, as the pressure gradient forcing is applied solely in the streamwise direction, which governs the dominant flow and transport.}
Figure \ref{fig:slab-averaged u}(a) shows the profile of plane-averaged superficial streamwise velocity $\spav{\overline{u}}$ and normalised by the friction velocity $u_\tau = \sqrt{{\tau_0}}$. 
Above the canopy region i.e., $z \geq h_{\max}$, the profile fits the logarithmic law
\begin{equation}
    \label{eq:loglaw}
    \spav{\overline{u}}(z) = \frac{u_\tau}{\kappa} \ln{\left( \frac{z-z_d}{z_0} \right)} \, ,
\end{equation}
where $\kappa$ is the Von Karman constant, $z_d$ is the displacement length and $z_0$ is the roughness length. 
Two different methods will be used to estimate the parameters $z_d$ and $z_0$. The first is to find the $z_d$ and $z_0$ directly from the profile in the inertial sublayer region (for example, here, we choose the fitted region above $2h_{\max}$ to the top end) that fits the logarithmic law best. This method takes $\kappa = 0.41$ to be given and results in $z_d = 40.0 \U m\approx 1.33h_m$ and $z_0 = 4.5 \U m \approx 0.15h_m$. The second method follows a hypothesis by \citet{Jackson1981}, that assumes the displacement height $z_d$ corresponds to the centre of pressure of the forces acting on the buildings, i.e.
\begin{equation}
    z_d = \frac{\int_0^{h_{\max}} z \spav{f_D} \d z}{\int_0^{h_{\max}} \spav{f_D} \d z} = \frac{1}{\tau_0} \int_0^{h_{\max}} \tau_{D} \d z\, ,
\end{equation}
where $\spav{f_D}$ is the distributed drag force and note that $\tau_D=\tau_{D;\infty}$ (see \S \ref{sec: steady_flow}). 
The displacement length calculated is $z_d = 35.1 \,\U m \approx 1.17 h_m$. In this approach, $\kappa$ is determined together with the roughness length $z_0$ to best fit the logarithmic law, with results $\kappa = 0.37$, and $z_0= 6.8 \,\U m \approx 0.23 h_m$. The velocity profiles fitted using both methods are plotted in Fig. \ref{fig:slab-averaged u}(a) as the dashed lines.
A prediction of $z_d$ and $z_0$ based on the plane index $\lambda_p$ \citep{Macdonald1998,Kanda2013} results values $z_d \approx 1.20 h_m , z_0 \approx 0.15 h_m$ for our case. All our results are generally close to this prediction with $z_0$ from the second method slightly higher. \review{As the displacement height $z_d$ and roughness length $z_0$ is typically related to the morphological parameters like plane area index $\lambda_p$, frontal area index $\lambda_f$, mean building height $h_m$ and the variance of the building height \citep{Grimmond1999} which are all predetermined for this case according to the London morphology profile, so that in general, we do not except a big variation in $z_d$ and $z_0$ among the random building-height distributions, although slight discrepancy would exist for example a distribution with a denser high-rise area may reduce the roughness length due to the sheltering effect \citep{Oke2017}.}

Figure \ref{fig:slab-averaged u}(b) shows the vertical evolution of the plane-averaged kinematic turbulent shear stress $-\spav{\overline{w^\prime u^\prime}}$ and dispersive stress $-\spav{\overline{w}^{\prime \prime} \overline{u}^{\prime \prime}}$. It shows that the turbulent shear stress $-\spav{\overline{w^\prime u^\prime}}$ increases with height to a peak near the maximum building height, and then decreases linearly with height as expected from classical boundary layer theory. 
The dispersive flux $-\spav{\overline{w}^{\prime \prime} \overline{u}^{\prime \prime}}$ is significant within the canopy with a peak near the mean building height. 
Above the canopy, the dispersive stress reduces to a very low value, which indicates that the upper limit of the roughness sublayer, or equivalently, the lower limit of the inertial sublayer, is at about $z=h_{\max}$. Note that this is a simple approximation; use of other quantities or definitions for the roughness sublayer, such as that of \citet{Placidi2015}, might work better for varying building heights but are beyond the scope of this study. The fact that the upper limit of the roughness sublayer is identical to the maximum building heights is quite unusual, since it implies that there is only an urban canopy layer but no roughness sublayer. Indeed, for uniform or random-height cube arrays \citep{Coceal2006, Xie2008, Sutzl2020, Lu2024}, the roughness sublayer extends above the building heights. This is likely the result of having a very large domain with a specific non-repeating height distribution of buildings.
There are 39 buildings out of the total 512 that have height $h_{\max}$, which is apparently insufficient to induce significant perturbations to the mean velocity field.

\begin{figure}
    \centering
    \includegraphics[width=14cm]{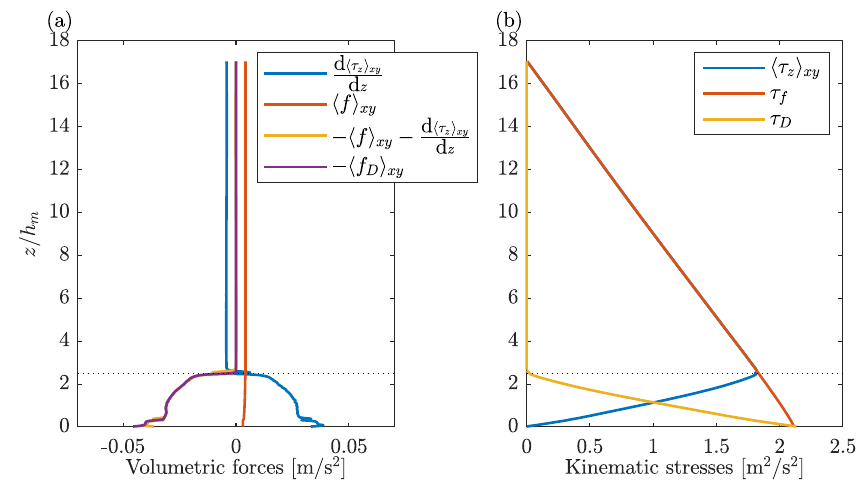}
    \caption{Plane-averaged momentum and cumulative stresses as a function of normalised height $z/h_m$. (a) Momenum balance Eq. \eqref{eq:NWP2}. (b) Cumulative kinematic stress balance Eq. \eqref{eq: int_force_balance2}. Dotted lines indicate the maximum building height.}
    \label{fig:mombalance}
\end{figure}

Figure \ref{fig:mombalance}(a) shows the plane-averaged force balance. 
The drag force $\spav{f_D}$ is determined in two ways: from the residual of the other two terms and from its direct definition Eq. \eqref{eq:slab_drag}. The good agreement between both shows that the momentum budget closes and is thus confirmation that the system is in a statistically steady state. 
Within the canopy region, the magnitude of the drag force decreases with height as expected because less space is occupied by building areas at larger heights. 
Above the canopy, the drag force vanishes and the pressure gradient matches the turbulent shear stress divergence.
Figure \ref{fig:mombalance}(b) illustrates the plane-averaged kinematic stress balance --- the integral version of the force balance Eq. \eqref{eq: int_force_balance2}. Above the canopy region, the stresses $\spav{\tau_z}$ and $\tau_{f}$ coincide in the absence of the drag force and both are linear with the height due to the constant gradient shown in Fig. \ref{fig:mombalance}(a).
Within the canopy region, the kinematic shear stress $\spav{\tau_z}$ increases roughly linearly with height, and the cumulative drag stress $\tau_{D}$ reduces from its maximal value at $z=0$ to zero at $h_{\max}$. 

\begin{figure}
    \centering
    \includegraphics[width=15cm]{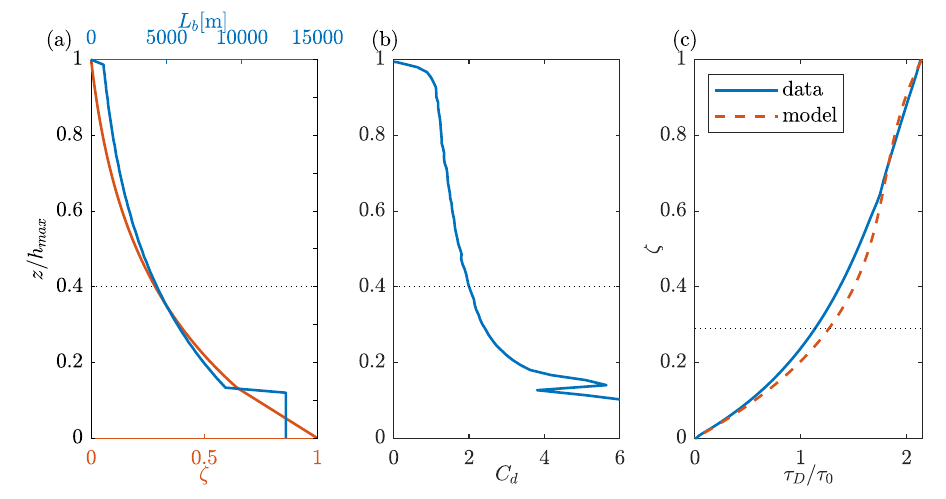}
    \caption{Distributed drag parameterisations as a function of scaled height $z/h_{\max}$. (a) Total frontal building width $L_b$ and scaled total frontal building area $\zeta$; (b) drag coefficient $C_d$ calculated from Eq. \eqref{eq: Cd}; (c) plane-averaged normalised cumulative drag stress $\tau_{D}/\tau_0$ against $\zeta$, overlaid with parameterisation Eq. \eqref{eq: tauD_zeta_para}. The dotted line indicates the location of the mean building height.} 
    \label{fig:zeta-tau relation}
\end{figure}

\review{We consider two methods in which the drag force can be parameterised.}
The first is to introduce a height-dependent drag coefficient $C_d(z)$ related to the forcing as \citep{Santiago2010}
\begin{equation}  \label{eq: Cd}
    \spav{f_D} =\frac{1}{2}  \frac{L_b}{A_T} C_d \spav{\overline{u}}^2 \, ,
\end{equation}
where we recall that $L_b(z)$ is the frontal width occupied by buildings, such that $\int_0^{h_{\max}} L_b(z) \, \d z  = A_F$. Note the absence of the fluid density in the expression above since $\spav{f_D}$ is a kinematic force. The second method is based on the observation that the cumulative drag $\tau_{D}(z)$ can be parameterised as \citep{Sutzl2020}
\begin{equation} \label{eq: tauD_zeta_para}
    \frac{\tau_{D}}{\tau_0} = 1.88 \zeta^3 -3.89 \zeta^2 + 3.01 \zeta,
\end{equation}
where $\zeta$ is the cumulative normalised frontal area, defined as
\begin{equation}
 \zeta(z) \equiv A_F^{-1}\int_z^{h_{\max}} L_b(z')\d z' \, .
\end{equation}
%
Figure \ref{fig:zeta-tau relation}(a) shows the vertical evolution of $L_b$ and $\zeta$ for the current urban geometry, and shows that $\zeta$ smoothly decreases from $\zeta=1$ at ground level to $\zeta=0$ at the maximum building height $h_{\max}$.

Figure \ref{fig:zeta-tau relation}(b) shows the drag coefficient $C_d$ calculated from Eq. \eqref{eq: Cd}. The coefficient is the largest near the ground due to very low velocities. The coefficient decreases from $4.0$ to $0.5$ over most of the building heights, i.e., $z/h_{\max} \geq 0.2$. The coefficient profile is generally consistent with the values reported in \citet{Coceal2004}. \review{The kink in the $C_d$ near $0.1<z/h_{\max}<0.15$ is due to the rapid decreases in building width $L_b$, because due to Eq. \eqref{eq: Cd} and drag and velocity profile, the production of $C_d$ and $L_b$ should decrease smoothly with the height.}
Figure \ref{fig:zeta-tau relation}(c) shows the relation between  $\tau_{D}$ and $\zeta$ together with Eq. \eqref{eq: tauD_zeta_para}. There is reasonably good agreement of the data with the parameterisation, although the current simulation does not show an inflection point near ground level ($\zeta=1)$.
This might suggest that Eq. \eqref{eq: tauD_zeta_para}, which was obtained from a best fit based on eight independent simulations, may need some further calibration, but this would need many simulations (e.g. using the \citet{Lu2024} dataset) to be done reliably. We will therefore keep using Eq. \eqref{eq: tauD_zeta_para} even though we could construct a better fitting relation.

\subsection{Multi-resolution analysis}
\begin{figure}
    \centering
    \includegraphics[width=14cm]{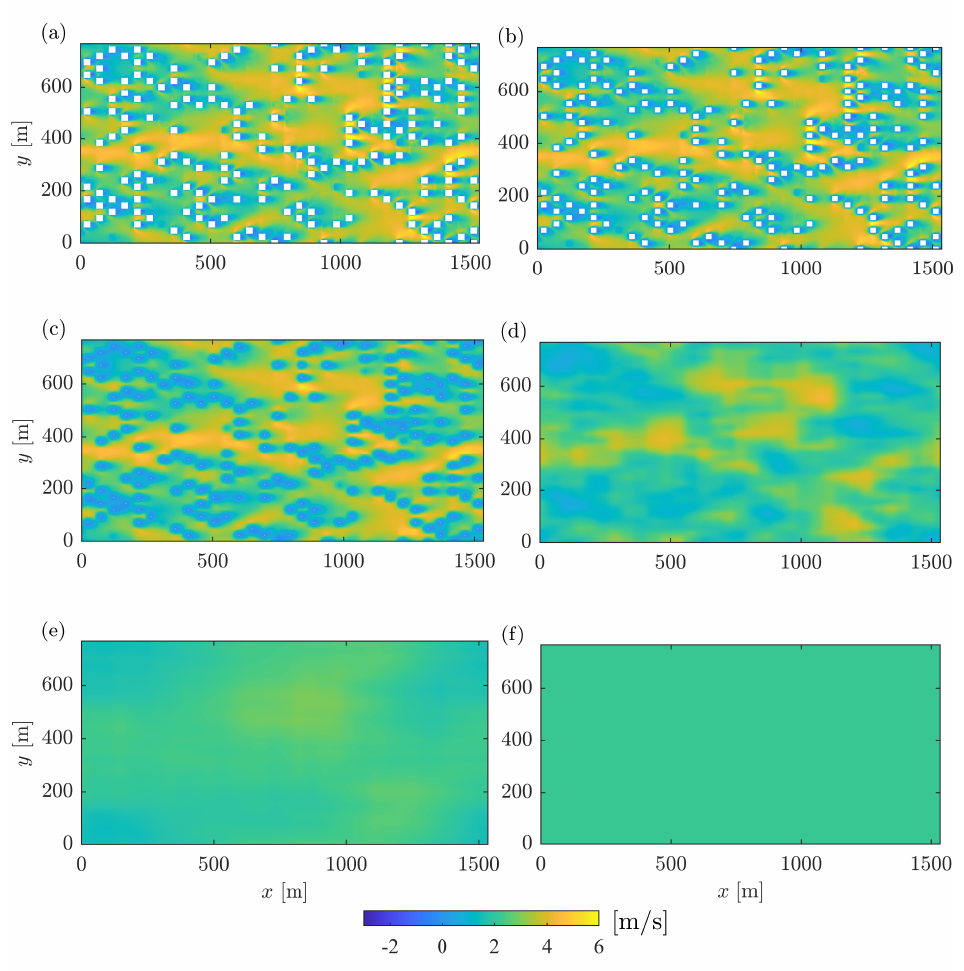}
    \caption{The streamwise velocity field $\overline u$ at various averaging lengths at the mean building level $z = 30$ m. (a) The original field $\overline{u}$, (b) $L =6$ m, (c) $L=24$ m, (d) $L= 96$ m, (e) $L= 384$ m, and (f) the plane-averaged field. The white boxes represent the buildings.}
    \label{fig:filtered fields}
\end{figure}

In this section, we will use the multi-resolution framework developed in \S \ref{sec:multiresolution}. 
To illustrate the coarse-graining, the mean streamwise velocity $\overline u$ at the mean building height $z=30$ m is shown in Fig. \ref{fig:filtered fields} at a number of different averaging lengths $L$.
Figure \ref{fig:filtered fields}(a) shows the velocity field at the original resolution. 
Recall that the resolution is $1.5$ m in both directions and that the white boxes represent the buildings where there is no fluid. 
Figure \ref{fig:filtered fields}(b-e) show the superficial velocity $\sav{\overline{u}}_L$ at averaging lengths $L$ of 6, 24, 96 and 384 m.
Figures \ref{fig:filtered fields}(b,c) show that the superficial averaging gradually fills in the areas in which no information about the flow is present.  
\review{Figure \ref{fig:filtered fields}(c) shows that the values of $\sav{\overline u}_{24}$ are unrealistically low at the original building locations}, as can be expected theoretically \citep{Whitaker1999, Schmid2019}.
At $L=96$ m however (Fig. \ref{fig:filtered fields}d), this no longer appears to be a significant effect.
At a resolution $L=384$ m (Fig. \ref{fig:filtered fields}e), the velocity $\sav{\overline u}_{384}$ is practically homogeneous.
Figure \ref{fig:filtered fields}(f) shows the plane-averaged value $\spav{\overline{u}}$ of the plane, which is homogeneous by definition.

\begin{figure}
    \centering
    \includegraphics[width=14cm]{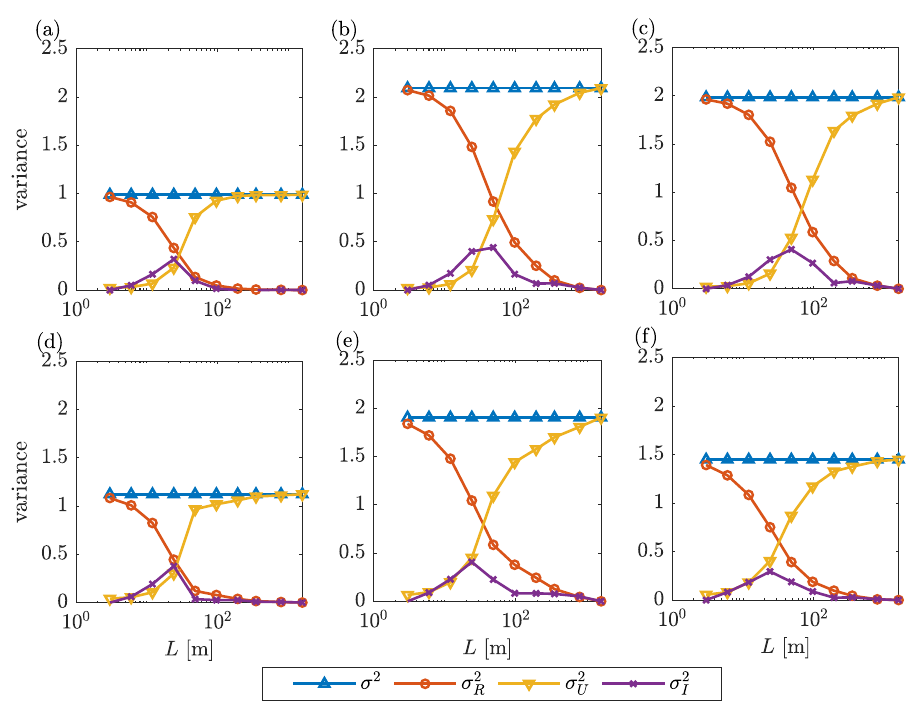}
    \caption{The total, resolved and unresolved variances of the streamwise velocity $\overline u$ (top panels) and the TKE $k$ (bottom panels) as a function of the averaging length $L$. The variances are taken at three different heights within the canopy region: (a,d) the pedestrian level $z/h_m=0.05$, (b,e) the mean building height $z/h_m=1.0$, (c,f) the canopy top $z/h_m=2.5$.} 
    \label{fig:variances}
\end{figure}

The variance of any quantity $\overline \varphi$ can be used to quantify its spatial heterogeneity  \citep[see also][]{Yu2023}.
\review{The plane-averaged variance of $\overline \varphi$ quantifies the deviation of the original field from its plane average,} which is defined as
\begin{equation}
    \sigma^2(z) \equiv \spav{(\overline \varphi(\vec x_\perp, z) - \spav{\overline \varphi})^2} \, ,
\end{equation}
with the understanding that $\spav{\cdot} = \sav{\cdot}_\infty$: this quantity represents the variance relative to the situation when the problem is fully homogenised in the $x$, $y$ directions (i.e. $L\gg \ell$). The total variance $\sigma^2$ can be decomposed as
\begin{equation}
  \label{eq:variancedecomp}
  \sigma^2 = \sigma^2_{R} + \sigma^2_{U} + \sigma^2_{I} \, ,
\end{equation}
\review{where the unresolved variance $\sigma^2_{U}$ quantifies the deviation of the original field from the spatial average, the resolved variance $\sigma^2_{R}$ quantifies the deviation of the spatial average from the plane average, 
and (twice) the interaction covariance ${\sigma^2_{I}}$ are defined as, respectively},
\begin{align*}
{\sigma^2_{U}} & \equiv {\spav{(\overline \varphi - \sav{\overline \varphi})^2}}, \\
{\sigma^2_{R}}& \equiv {\spav{(\sav{\overline \varphi} - \spav{\overline \varphi})^2}}, \\
{\sigma^2_{I}} & \equiv {2 \spav{(\overline \varphi - \sav{\overline \varphi})(\sav{\overline \varphi} - \spav{\overline \varphi})}}.
\end{align*}
These variances are dependent on the height $z$ and the averaging length $L$.

Figure \ref{fig:variances} shows the plane-averaged variances of streamwise velocity $\overline u$ and TKE $k$ as a function of the averaging length $L$ at three levels within the canopy region, specifically, at the pedestrian level, at the mean height and at the maximum building height. 
The velocity and TKE variance data have very similar trends.  
By definition, the total variance $\sigma^2$ is independent of the averaging length $L$. 
\review{Generally, for both quantities, the variances are small near the ground, are larger at the mean height and the maximum height levels due to the building height distribution, and should gradually vanish above the canopy region as the flow becomes uniform.}

The dependence of the variances on the averaging length $L$ is similar for all heights and both quantities. 
The resolved variances $\sigma^2_{R}$ decrease while the unresolved variances $\sigma^2_{U}$ increase with increasing $L$. This is consistent with Fig. \ref{fig:filtered fields}, which shows inhomogeneity (large resolved variance) at low values of $L$ and homogeneity (low resolved variance) at high values of $L$, explaining the resolved variance. 

The interaction term $\sigma^2_{I}$ is always negligible at very high or low values of $L$ as either $\sigma^2_{R}$ or $\sigma^2_{U}$ would dominate in Eq. \eqref{eq:variancedecomp}. Indeed, for very high resolutions, most variance is resolved, i.e., it is not necessary to model the unresolved part; while for very low resolutions, the variance is largely unresolved. The interaction term thus provides information on the scales at which resolved and unresolved scales interact most strongly, and thus provides information on the characteristic urban length scale $\ell$ introduced in \S \ref{sec:multiresolution}. For all figures, $\sigma^2_{I}$ peaks around $L \approx 24-96$ m, and therefore $\ell \approx 50$ m for this case, which is about twice the building width $W$. Note that this value is consistent with the averaging length at which the resolved variance $\sigma^2_{R}$ and unresolved variance $\sigma^2_{U}$ are equal to each other, which would be an alternative method to infer a characteristic urban lengthscale $\ell$. For averaging lengths $L>200$ m, most variance is unresolved and homogeneity can be assumed. 

\begin{figure}
    \centering
    \includegraphics[width=10cm]{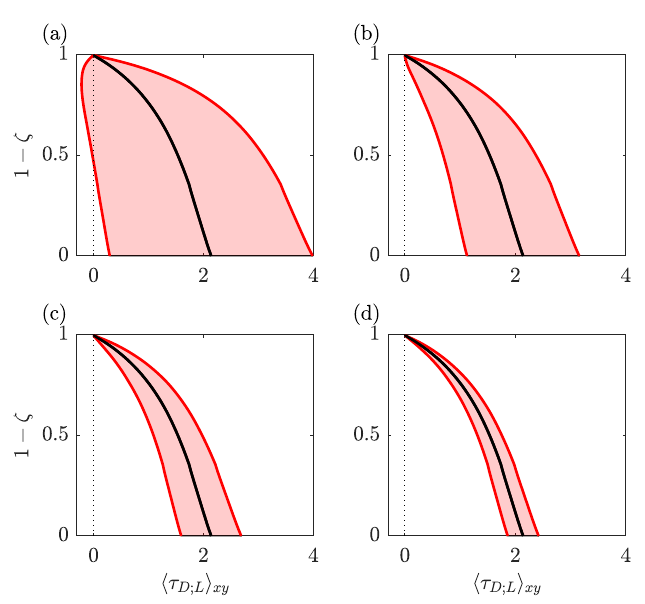}
    \caption{Vertical profile of plane-averaged cumulative drag stress $\spav{\tau_{D; L}}$ in the canopy layer against the normalised frontal area $\zeta$. (a) $L =96$ m. (b) $L=192$ m. (c) $L=384$ m. (d) $L=786 $ m, The standard deviation $\sigma_R$ is marked by the red band.}
    \label{fig:tau_D}
\end{figure}

\subsection{Drag distribution}

As discussed in \S \ref{sec: Re-u-momentum equation}, the drag distribution $\sav{f_D}$ or equivalently the cumulative kinematic shear stress $\sav{\tau_D}$ is a key quantity that requires parameterisation for non-building resolving NWP simulations. 
Figure \ref{fig:tau_D} shows the plane-averaged value of the cumulative drag force $\spav{\tau_{D; L}}$ at different averaging lengths $L$ together with its standard deviation $\sigma_R$ as a function of the scaled height $1-\zeta$.
Here, we have chosen $1-\zeta$ over $\zeta$ so the canopy top is presented at the top of the figures.
Note that $\spav{\tau_{D; L}}$ is independent of $L$, as shown in the black line.
The variances are typically small near the canopy top and increase with decreasing height. 
It is also clearly visible that the spread in $\tau_{D; L}$  increases as the averaging length $L$ decreases.
This suggests that the plane-averaged drag is not representative of the local value at a specific height. 

\review{In the following, we perform an \emph{a priori} analysis  at different averaging lengths $L$ and test the performance of the \citet{Sutzl2020} parameterisation Eq. \eqref{eq: tauD_zeta_para}. 
} 
In order to do this, we need to make the parameterisation local, which can be done as follows:
\begin{equation} \label{eq:3D-para}
    \frac{\tau_{D; L}}{\tau_{0;L}} = 1.88 \zeta_L^3 -3.89  \zeta_L ^2 + 3.01  \zeta_L \, ,
\end{equation}
where $\tau_{0;L} \equiv \tau_{D;L}(0) = \tau_{f;L}(0)$ is the superficially averaged kinematic surface stress and 
\begin{equation}
\zeta_L(\vec x_\perp, z) \equiv \frac{1}{\lambda_{f;L}}
\int_z^h \volav {\rho_{L}}_L(\vec x_\perp, z') \, \d z' \, ,
\end{equation}
is the scale-dependent cumulative normalised frontal area, where $\rho_L$ is the frontal area density field described in \S \ref{sec:implementation} (note that $L_{b;L} = L^2 \sav{\rho_{L}}_L$). In the above equation, the scale-dependent frontal area index $\lambda_{f;L}$ is defined as
\begin{equation}
\lambda_{f;L}(\vec x_\perp) \equiv \int_0^h \volav {\rho_{L}}_L(\vec x_\perp, z') \, \d z' \, .
\end{equation}
In the limit of $L\gg \ell$, we recover the classical quantities $\tau_0 \equiv \tau_{0;\infty}$ and $\lambda_f \equiv \lambda_{f;\infty}$.
\begin{figure}
    \centering
    \includegraphics{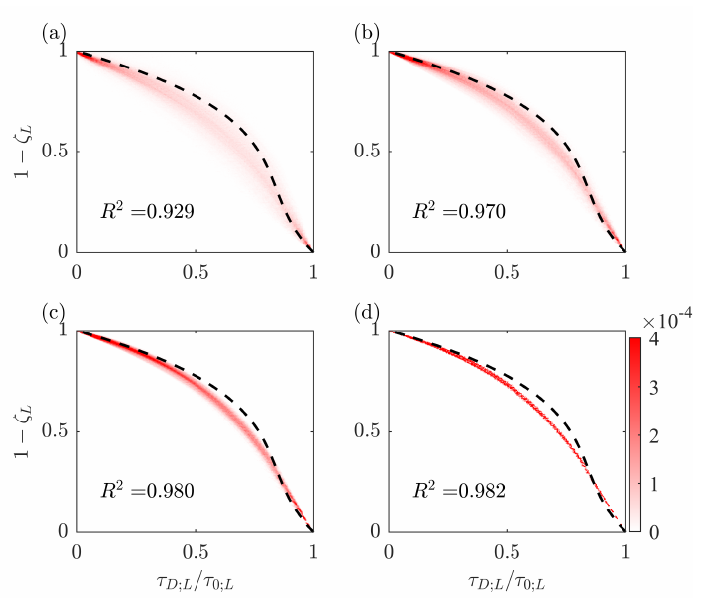}
    \caption{Normalised local cumulative drag stress $\tau_{D; L}/\tau_{0;L}$ against the local scaled frontal area $\zeta_L$ at different averaging lengths. \review{The colour represents the statistical probability density function value of the data appearing at each location.} (a) $L =96$ m, (b) $L=192$ m, (c) $L= 384$ m, (d) $L= 786$ m. The dashed line represents the parametrisation Eq. \eqref{eq:3D-para}. The coefficient of determination $R^2$ is labelled to show the \emph{a priori} performance of parametrisation.}
    \label{fig:tau_D scatter}
\end{figure}

Figure \ref{fig:tau_D scatter} shows the \review{probability density function (PDF) value of the distribution of the local cumulative drag stress $\tau_{D; L}$ against the local scaled frontal area $\zeta_L$, at different resolutions and overlaid with the parametrisation Eq. \eqref{eq:3D-para}.} 
Figures \ref{fig:tau_D scatter}(c,d) show that at averaging lengths $L=384$ and $786$ m, there is very little scatter in the data as the fields are relatively homogeneous and the parametrisation works well with a very high coefficient of determination. 
This clearly shows that accurate parameterisation requires local quantities rather than plane-averaged quantities (Fig. \ref{fig:tau_D}).

As the resolution increases, the scatter increases due to heterogeneity effects.
Indeed, Eq. \eqref{eq:heterogeneityeffects} shows that, whenever there are heterogeneity effects, the horizontal terms in the momentum equation become important, which implies that the vertical terms are no longer in balance. The latter is, of course, what Eq. \eqref{eq:3D-para} is built on. However, a high coefficient of determination $R^2$ illustrates that the parameterisation still fits the data reasonably well at $L=96$ m (Fig. \ref{fig:tau_D scatter}a), \review{indicating that the vertical terms still dominate}.
However, at averaging lengths below 96 m, the agreement between the data and the parameterisation quickly deteriorates, we find that the coefficient of determination dramatically reduces to $0.23$ at $L = 48$ m, which means a poor match, \review{indicating that further decreasing $L$, the horizontal terms can no longer be ignored}. Indeed, below $L = 96$ m, there is a significant increase in the resolved variance (Fig. \ref{fig:variances}), and thus the resolved heterogeneity. Note that, the data distribution within Fig. \ref{fig:tau_D scatter} is much narrower than that shown in Fig. \ref{fig:tau_D} which is associated with the same quantity $\tau_{D; L}$, the reason is that buildings with different heights would accumulate various of drag stress from their bottom to top, therefore, Fig. \ref{fig:tau_D} shows a wide deviation. However, in Fig. \ref{fig:tau_D scatter}, normalising the cumulative stress by its local value greatly reduces the spread of the data.

\begin{figure}
    \centering
    \includegraphics{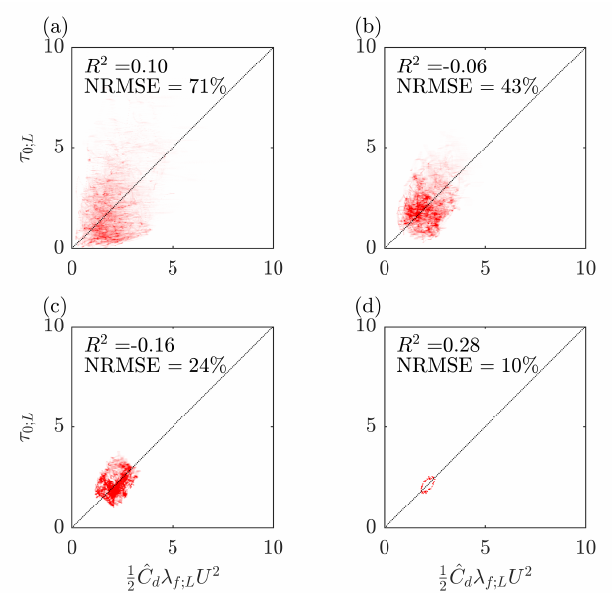}
    \caption{Comparison between the cumulative drag $\tau_{0; L}$ and the shear stress parameterisation Eq. \eqref{eq:linear tau_D0} at various averaging lenghts. (a) $L =96$ m, (b) $L=192$ m, (c) $L= 384$ m, (d) $L= 786$ m. The dotted line indicates perfect agreement. The coefficient of determination $R^2$ is labelled to show the performance of the parametrisation.}
    \label{fig:tau_D,0}
\end{figure}

A potential drawback of Eq. \eqref{eq:3D-para} is that, even though it agrees with the data well at averaging lengths larger than 96 m, it does not provide the information for the superficially averaged local shear stress  $\tau_{0; L}$.
Here we model $\tau_{0; L}$ using a single drag coefficient $\hat{C}_d$ as \citep{Coceal2004, Belcher2005, Sutzl2021}
\begin{equation} \label{eq:linear tau_D0}
    \tau_{0; L} = \frac{1}{2} \hat{C}_d \lambda_{f;L} U^2 \, ,
\end{equation}
where $U = \sav{\overline{u}}_L(\vec x_\perp, z=30)$ is the reference (superficial) wind velocity at $30$ m which coincides with the mean building height. 
Note that $\hat{C}_d$ is distinct from  $C_d$ in Fig. \ref{fig:zeta-tau relation}, which is height-dependent.
Using the plane-averaged data, the appropriate value for $\hat{C}_d$ in the parameterisation above is $\hat{C}_d = 2.49$.

In Fig. \ref{fig:tau_D,0} we show the distribution of $\tau_{0; L}$ against the parametrisation $\frac{1}{2} \hat{C}_d \lambda_{f;L} U^2$. For $L=768$ m (Fig. \ref{fig:tau_D,0}d), the data has a very small spread and has the right value, as expected since $\hat{C}_d$ was inferred from the plane-averaged data.
As $L$ becomes smaller, the spread of the data becomes wider, although still centred around the diagonal.
For $L=96$ m (Fig. \ref{fig:tau_D,0}a), the parametrisation does not predict the local surface stress well. However, the coefficients of determination $R^2$ of all four figures are very poor, which is because $R^2$ becomes less informative when the data show little variation or are very close to their mean value, even for good modellings. With this regard, we also evaluate the normalised root-mean-square error (NRMSE) to identify the \emph{a priori} performance of the parametrisation --- the lower the NRMSE is, the better the model predicts. This error is generally consistent with the data distribution in the figure.

\subsection{Canopy turbulence closures}
As discussed in \S \ref{sec: Re-u-momentum equation}, the other key process that requires parameterisation for non-building resolving NWP simulation is the unresolved stress $\sav{\tau_i}$. 
Consistent with (\ref{eq:NWP},\ref{eq: tau_i}), we consider the vertical flux $\sav{\tau_z} = \sav{\overline{w}} \sav{\overline{u}} - \sav{\overline{w} \, \overline{u}} - \sav{\overline{w^\prime u^\prime}}$, which comprises a turbulent momentum flux $\sav{\overline{w^\prime u^\prime}}$ and a residual term $\sav{\overline{w}} \sav{\overline{u}} - \sav{\overline{w} \, \overline{u}}$. As demonstrated in Eq. \eqref{eq: decomp_wphi}, this residual term can be further decomposed associated, but for simplicity, we regard the residual as a whole and name it a dispersive-like term.

A common way to represent the unresolved fluxes in the canopy region is based on $K$ theory, which parameterises the flux by linking the flux to the strain rate tensor $S_{ij}$ employing an eddy diffusivity $K_m$ \citep{pope_2000, Wyngaard2010}. For the vertical turbulence flux of horizontal momentum, this results in
\begin{equation} \label{eq: K-theory}
    \volav{\overline{w^\prime u^\prime}} = - 2 K_m S_{31} \, , \quad \textnormal{where}\quad S_{ij} \equiv \frac{1}{2} \left(\frac{\partial \volav{\overline{u}_i}}{\partial x_j} + \frac{\partial \volav{\overline{u}_j}}{\partial x_i} \right) \, .
\end{equation}
The eddy-diffusivity coefficient $K_m$ can be modelled in many different ways \citep{Hanjalic2022}. We investigate two models here, the $k-l$ model \citep{Deardorff1972, Bougeault1989} and the $k-\omega$ model \citep{Wilcox1998}:
\begin{equation} \label{eq: turbulence model}
    K_m = C_k l_k \sav{{k}}^{1/2} \quad \textnormal{and} \quad K_m = \frac{\sav{k}}{\omega} \, ,
\end{equation}
respectively, where $C_k$ is a model constant, $l_k$ is a turbulent mixing length, $\sav{{k}}$ is the superficially averaged turbulence kinetic energy and $\omega$ is an inverse turbulence time scale.

\begin{figure}
    \centering
    \includegraphics[width=15cm]{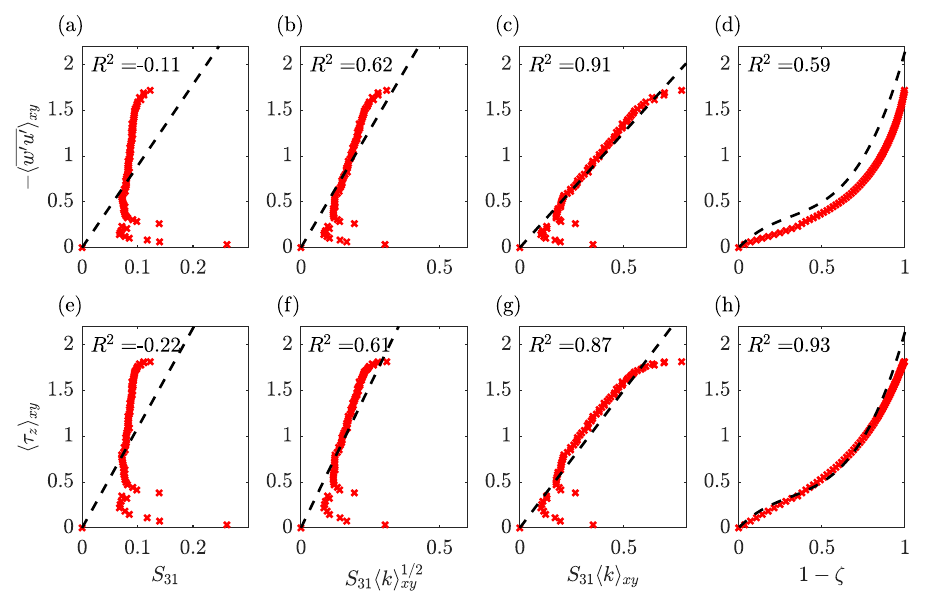}
    \caption{Suitability of various turbulence closures to parameterise the turbulence shear stress $-\spav{\overline{w ^\prime u^ \prime }}$ (top row) and the total unresolved stress $\spav{\tau_z}$ (bottom row). (a,e) Constant eddy viscosity, (b,f) $k-l$ model, (c,g) $k-\omega$ model, and (d,h) parameterisation Eq. \eqref{eq:3D-para}. Agreement of the data with the dashed lines implies agreement with the model. The data are shown within the canopy region. The coefficient of determination $R^2$ is labelled to show the \emph{a priori} performance of parametrisation.}
    \label{fig:nuk}
\end{figure}

First, we will study the closures for plane-averaged statistics. 
Figures \ref{fig:nuk}(a-c) show the turbulence fluxes $\spav{\overline{w^\prime u^\prime}}$ in the canopy layer with the x-axes chosen such that a straight line starting at the origin indicates agreement with the closure.
Figure \ref{fig:nuk}(a) shows that the turbulence flux in the canopy region is practically independent of $S_{31}$, suggesting that the eddy diffusivity $K_m$ cannot be assumed constant in the canopy layer. The best-fit value (associated with the dashed line is $K_m= 8.91$ m$^2$/s, which is greater than the value of $1$ m$^2$/s reported by \citet{Lu2024}.  This is likely due to the difference in geometry. However, the negative coefficient of determination $R^2$ indicates that the line fitting from the origin shows a poor match.
Figure \ref{fig:nuk}(b) shows that the mixing length model performs better, but although the data is pretty much on a straight line it does not pass through the origin. The best-fit results in $l_k=13.00$ m, here we have assumed the model constant $C_k = 0.4$ \citep{Bougeault1989}. The ratio $C_kl_k/h_{\max}=0.07$ is comparable with \citep{Nazarian2020} at the same urban density $\lambda_p$.
Only for the $k-\omega$ model (Fig. \ref{fig:nuk}c) does the data compare well with a line through the origin (with $R^2$ very close to $1$). The best-fit returns a value $\omega=0.40$ s$^{-1}$. 

Figures \ref{fig:nuk}(e-g) present the same figures, but now using the kinematic unresolved shear stress $\spav{\tau_z}$ which includes the dispersive-like term.
The small difference between Figures \ref{fig:nuk}(a-c) and (e-g) is due to the fact that the turbulence fluxes dominate everywhere in the canopy layer (Fig. \ref{fig:slab-averaged u}b).
All the figures show that considering the dispersive-like terms slightly reduces the models' performance. However, again, the $k-\omega$ model can be seen to perform the best with the best-fit values $\omega=0.33$ s$^{-1}$. 

Another way to parameterise $\volav{\tau_z}$ without having to rely on auxiliary variables like $k$ is to use the drag parametrisation Eq. \eqref{eq:3D-para}. Invoking the constant stress layer assumption often used in boundary layer theory \citep{Schlichting2017} implies that  $\tau_f\approx\tau_0$ in the canopy layer, which after substitution into Eq. \eqref{eq: int_force_balance2} and making use of Eq. \eqref{eq:3D-para} immediately results in 
\begin{equation} \label{eq: tau_z-zeta}
    \frac{\spav{\tau_z}}{\tau_{0}} = 1 - \frac{\tau_D}{\tau_{0}} = 1 - \zeta^3 -3.89 \zeta^2 + 3.01 \zeta \, .
\end{equation}
The distribution of $\spav{\overline{w^\prime u^\prime}}$ and $\spav{\tau_z}$ against $\zeta$ are shown in figures \ref{fig:nuk}(d,h), respectively. Since the parameterisation is derived for $\spav{\tau_z}$, it always overestimates the $\spav{\overline{w^\prime u^\prime}}$ by definition --- the total stress is in better agreement with the parameterisation than the turbulence flux only (Fig. \ref{fig:nuk}h). Near the top, for $1 - \zeta > 0.9$, the parametrisation slightly overestimates the total stress. Comparing the bottom row, the parametrisations in Fig. \ref{fig:nuk}(g,h) of total unresolved stress have the largest $R^2$ to fit the data.

\begin{figure}
    \centering
    \includegraphics[width=15cm]{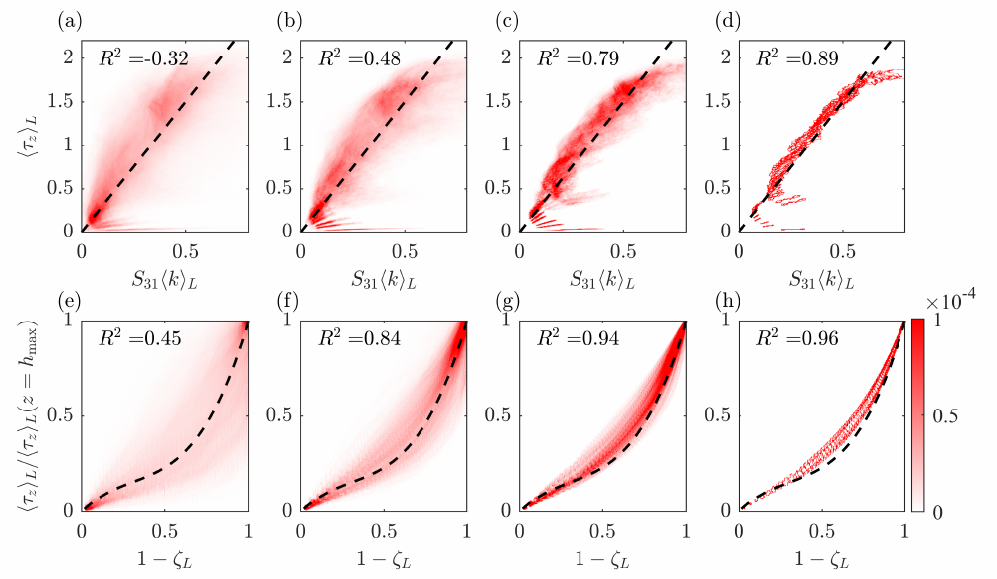}
    \caption{Skill of the $k-\omega$ model (top row) and Eq. \eqref{eq:3D-para} (bottom row) to parameterise the total unresolved vertical stress $\sav{\tau_z}_L$ at various averaging lengths. (a,e) $L = 96$ m, (b,f) $L = 192$ m, (c,g) $L = 384$ m, (d,h) $L = 768$ m. Overlaid with the parametrisation from Fig. \ref{fig:nuk}(g,h) for the top and bottom row, respectively. The coefficient of determination $R^2$ is labelled to show the \emph{a priori} performance of parametrisation.}
    \label{fig:uw-komega}
\end{figure}

Figure \ref{fig:uw-komega} shows the two best-performing parameterisations of total unresolved stress at different averaging lengths $L$. The top panels show the $k-\omega$ model predictions for the total unresolved shear stress $\sav{\tau_z}_L$ and the bottom row shows the predictions derived from the drag parameterisation Eq. \eqref{eq: tau_z-zeta}. Note that the stresses in the bottom row are normalised by $\sav{\tau_z}_L$ at maximum building height $z = h_{\max}$ in order to converge to the unit. However, under the constant stress layer assumption \citep{Schlichting2017}, $\sav{\tau_z}_L(z = h_{\max}) \approx \tau_0$ which is consistent with the parameterisation Eq. \eqref{eq: tau_z-zeta}.
In order to create a scale-dependent parameterisation for Eq. \eqref{eq: tau_z-zeta}, simply replace $\spav{\tau_z}$ with $\sav{\tau_z}_L$, $\tau_0$ with $\tau_{0;L}$, and $\zeta$ with $\zeta_L$.
As expected, as the averaging length reduces, the data spread around the diagnostic line increases with the reduced coefficient of determination $R^2$. However, $R^2$ is larger in the bottom row than the top row at the same resolution, suggesting that the parametrisation Eq. \eqref{eq: tau_z-zeta} has higher fidelity than the $k-\omega$ closure.
We also note that, compared with the drag parametrisation (Fig. \ref{fig:tau_D scatter}), the coefficients $R^2$ of stress parametrisation Eq. \eqref{eq: tau_z-zeta} are lower at the same resolution, and reduce more rapidly with the reducing resolution. This might be because of the assumption of constant stress.

\section{Conclusions}
\review{A multi-scale framework was developed for investigating the flows over heterogeneous urban areas.
At the core of the framework is a coarse-graining method that employs convolution filters to obtain coarse-grained fields from a building-resolved high-resolution simulation.
%
By employing Fast-Fourier transforms, a computationally efficient procedure was developed to carry out coarse-graining, which is particularly important for the analysis of high-resolution simulations as considered here. 
}

One of the central questions the paper aimed to address was: "What are the requirements for non-building-resolving NWP models as their resolution is increased?" 
To answer this question, a quasi-realistic case was considered, inspired by the building height distribution of London.
The multi-scale framework revealed that the flow has a characteristic length scale $\ell\approx 50$ m, which is the resolution at which as much variance is resolved as is unresolved. This corresponds to roughly twice the building width. Heterogeneity effects, which challenge the NWP parameterisations for the land surface, become significant at resolutions of about 200 m \review{and below} for the case under consideration.

Given that future NWP models will likely run at resolutions at which heterogeneity effects are important, an immediate follow-up question is how to incorporate heterogeneity.
This requires a land-surface model that resolves the vertical distribution of the flow and turbulence \citep{Lean2024} and requires: i) a model for the distribution of the drag over the canopy; and ii) a model for the unresolved turbulence and dispersive stresses in the canopy.
Since NWP models do not resolve the urban morphology and it is undesirable to have to modify the governing equations inside the canopy (e.g. the mass conservation equation), superficial (comprehensive) averages are the appropriate formulation to consider, see Eq. \eqref{eq:NWP}.
It is therefore highly desirable that the drag distribution and unresolved stress parameterisation are formulated in terms of superficial quantities also.

It was shown that the distributed drag parameterisation of \citet{Sutzl2020}, which was obtained from plane-averaged analysis of an ensemble of eight different urban landscapes, can be successfully extended to a local formulation, cf Eq. \eqref{eq:3D-para}.
The parameterisation generalises excellently, with an $R^2$ value of 0.929 at 96 m resolution. 
The attractive feature of this parameterisation is that it only requires the local distributed frontal area $\zeta_L$ which can be calculated \emph{a priori} and can be straightforwardly incorporated in NWP models \citep{Sutzl2021}.
Although this parameterisation is excellent for predicting the local drag distribution, the actual drag requires knowledge of the local surface stress which remains challenging and requires further work; a local version of the \cite{Coceal2004, Belcher2005} parameterisation Eq. \eqref{eq:linear tau_D0} has a normalised root-mean-square error of 71\% at 100 m resolution.


Several closures were considered for the unresolved and turbulent stresses in the urban canopy layer.
The two best-performing models were the $k-\omega$ model, which requires information about the turbulence kinetic energy $k$ and the inverse turbulence timescale $\omega$, and a new stress parameterisation based on the distributed drag, cf Eq. \eqref{eq: tau_z-zeta}, that does not require further parameters. 
At a resolution of 96 m, the latter performs better than the former, although both parameterisations have a substantial scatter associated with local heterogeneity effects. 

The results reported here indicate that it is possible to incorporate substantial heterogeneity effects in non-building-resolving NWP models with relative ease.
However, further work is required.
First, it is desirable to ensure the validity of the parameterisations Eq. \eqref{eq: tauD_zeta_para} and Eq. \eqref{eq: tau_z-zeta} on a much larger flow database, e.g. that of \cite{Lu2024}. 
Furthermore, the atmosphere is non-neutral most of the time, and distributed versions of the sensible and latent heat fluxes, including the interaction with the surface stress, need to be explored. 

\section*{Acknowledgements}
The support of the ARCHER2 UK National Supercomputing Service (project ARCHER2-eCSE05-3) and the NERC highlight grant ASSURE: Across-Scale ProcesseS in Urban Environment (NE/W002868/1) is acknowledged. 

\section*{Data availability}
The data that support the findings of this study, as well as the coarsegraining routines are openly available in Zenodo at \citet{vanReeuwijk_Huang_2025}.

\section*{Declaration of interests}
The authors report no conflict of interest.

\appendix

\section{Volume-averaging with convolution filters}
\label{sec:volumeaveraging}

The \citet{Whitaker1999} volume averaging approach can be cast into a convolution filter approach by introducing a kernel $\mathcal G$ to specify the averaging volume, which is a function that is $1/V$ within the averaging domain and 0 outside of it, where $V$ is the total averaging volume. 
Using this kernel, the superficial volume average of an arbitrary scalar  $\varphi(\vec x)$ is given by
\begin{equation}
\label{eq:volav}
 \volav{\varphi}(\vec x) \equiv \int_{\Omega_f(\vec x)} \mathcal G(\vec x - \vec y) \varphi(\vec y) \d \vec y \, ,
\end{equation}
where $\vec x$ is the 3-D location and $\int \mathcal G (\vec x) \d \vec x = 1$, i.e.\ the kernel is normalised. 
Note that in Eq. \eqref{eq:volav}, integration only takes place over the fluid phase. 
The fluid volume fraction $\varepsilon = V_f/V$, where $V_f$ is the volume occupied by fluid inside the averaging volume, is given by
\begin{equation}
  \varepsilon(\vec x) = \int_{\Omega_f(\vec x)} \mathcal G(\vec x - \vec y) \d \vec y.
\end{equation}

\subsection{Volume to area integration}
An area-averaging filter can be obtained by defining $\mathcal G$ as
\begin{equation}
\label{eq:Garea}
\mathcal G(\vec x) = \mathcal A(\vec x_\perp) \delta(z) \, ,
\end{equation}
where $\delta$ is the Kronecker delta and  $\vec x = [\vec x_\perp, z]^T$.
With this definition of $\mathcal G$, the volume integral will become an area integral. The planar filter is also normalised:
\begin{equation*}
\int \mathcal{G}(\vec x) \d \vec x = \int \int \mathcal{A}(\vec x_\perp) \delta(z) \d \vec x_\perp \d z = 
\int \mathcal{A}(\vec x_\perp) \d \vec x_\perp  = 1 \, .
\end{equation*}
Using Eq. \eqref{eq:Garea}, the superficial average Eq. \eqref{eq:volav} becomes
\begin{equation}
 \label{eq:areaav_app}
 \volav{\varphi}(\vec x) =  \int_{\Omega_f(\vec x)} \mathcal A(\vec x_\perp - \vec y_\perp) \varphi(\vec y_\perp, z) \d \vec y_\perp \, .
\end{equation}
The surface fluid fraction $\varepsilon = A_f/A$, where $A_f$ is the fluid area and $A$ is the filter area, is given by
\begin{equation} \label{eq: fluid index}
  \varepsilon(\vec x) = \int_{\Omega_f(\vec x)} \mathcal A(\vec x_\perp - \vec y_\perp) \d \vec y_\perp \, ,
\end{equation}

\subsection{Differentiation rules}

We start with deriving the differentiation rule for the divergence. Note that 
\begin{equation}
  \label{eq:div1}
  \sav{\nabla \cdot  \vec F}
  = 
  \int_{\Omega_f} \mathcal A \nabla \cdot  \vec F \d \vec y_\perp 
  = 
    \int_{\Omega_f} \nabla \cdot (\mathcal A \vec F) \d \vec y_\perp
  - \int_{\Omega_f} \vec F \cdot \nabla \mathcal A  \d \vec y_\perp
\end{equation}
Commuting the integration and differentiation operators in the first term needs to be done with care, since $\Omega_f$ is dependent on space. It requires the 
identity \citep[see][for a derivation]{Maarten_2021}:
\begin{equation}
  \label{eq:vecid}
  \int_{\Omega_f} \nabla \cdot  \vec G \d \vec y_\perp
  =
  \frac{\partial }{\partial z} 
  \int_{\Omega_f} G_z \d \vec y_\perp
  - \oint_{\partial \Omega_f}   \frac{\vec G \cdot \vec N}{|\vec N_\perp|} \d s.
\end{equation}
Using Eq. \eqref{eq:vecid}, Eq. \eqref{eq:div1} becomes
\begin{equation}
  \label{eq:divtheorem1}
  \sav{\nabla \cdot  \vec F}
 = 
 \frac{\partial \sav{F_z}}{\partial z} 
  - \int_{\Omega_f} \vec F_\perp \cdot \nabla_\perp \mathcal A  \d \vec y_\perp
  - \oint_{\partial \Omega_f} \mathcal A  \frac{\vec F \cdot \vec N}{|\vec N_\perp|} \d s,
\end{equation}
where use was made of the fact that $\partial \mathcal A / \partial z = 0$.
The first term in this expression is the vertical gradient of the superficial average of $F_z$.
The second term represents the filtered horizontal gradient (see below). 
The last term represents the fluxes that are exchanged between the fluid and the solid domain inside the averaging area. 
Note that \citep{Leonard1974, pope_2000}
\begin{equation*}
\begin{split}
  \int_{\Omega_f} \vec F_\perp(\vec y_\perp) \cdot \nabla_\perp \mathcal A(\vec x_\perp - \vec y_\perp)  \d \vec y_\perp
  & =\int_{\Omega_f} \vec F_\perp(\vec y_\perp) \cdot  \frac{\partial}{\partial \vec y_\perp} \mathcal A(\vec x_\perp - \vec y_\perp)  \d \vec y_\perp \\
  &= 
  - \int_{\Omega_f} \vec F_\perp(\vec y_\perp) \cdot \frac{\partial }{\partial \vec x_\perp} \mathcal A(\vec x_\perp 
  - \vec y_\perp)  \d \vec y_\perp \\
  &= 
  - \frac{\partial }{\partial \vec x_\perp} \cdot \int_{\Omega_f} \vec F_\perp(\vec y_\perp)  \mathcal A(\vec x_\perp 
  - \vec y_\perp)  \d \vec y_\perp \\
  &= -\nabla_\perp \cdot \sav{\vec F_\perp},
\end{split}
\end{equation*}
which implies that the vector form of the spatial averaging theorem takes the form
\begin{equation}
  \label{eq:divtheorem}
  \sav{\nabla \cdot  \vec F}
 = 
 \nabla \cdot  \sav{\vec F}  - \oint_{\partial \Omega_f} \mathcal A  \frac{\vec F \cdot \vec N}{|\vec N_\perp|} \d s. 
\end{equation}
Substitution of $\vec F = \varphi \vec e_i$ where $i \in \{x,y,z\}$  into Eq. \eqref{eq:divtheorem} results in the the spatial averaging theorem for scalars:
\begin{equation}
  \label{eq:gradtheorem}
  \volav{\nabla \varphi}
  = \nabla \sav{\varphi}
   - \oint_{\partial \Omega_f} \mathcal A \varphi \frac{\vec N}{|\vec N_\perp|} \d s \, .
\end{equation}

\bibliographystyle{spbasic_updated}     
\bibliography{References.bib}

\end{document}